\documentstyle[a4,epic,eepic]{article}
\topmargin - 1.0 true cm
\def\bsll{B\to X_s l^+l^-}
\def\absll{{\bar{B}\to X_s l^+l^-}}
\def\bsg{B\to X_s \gamma}
\def\beq{\begin{equation}}
\def\eeq{\end{equation}}
\def\bea{\begin{eqnarray}}
\def\eea{\end{eqnarray}}
\def\nn{\nonumber}

\def\LLop{\bar{s}_L \gamma_{\mu} b_L \bar{l}_L \gamma^{\mu} l_L}
\def\LRop{\bar{s}_L \gamma_{\mu} b_L \bar{l}_R \gamma^{\mu} l_R}

\begin{document}
\begin{flushright}
{\bf hep-ph/0102041}\\
HUPD-0014
\end{flushright}
\bigskip
\begin{center}
{\large
{\Large CP Asymmetry of $\bsll$ in Low Invariant Mass Region}\\
\bigskip
{S. Fukae$^a$\footnote{fukae@hiroshima-u.ac.jp},\\
{$^a$\it Department of Physical Science, Hiroshima University},\\ {\it Higashi
Hiroshima 739-8526, Japan}}
}
\end{center}
\bigskip
\begin{abstract}
I analyzed the CP asymmetry of $\bsll$ based on model-independent analysis which
includes twelve independent four Fermi operators. The CP asymmetry is suppressed
in the Standard Model, however, if some new physics make it much larger, the
present or the next generation B factories may catch the CP violation in this
decay mode. In this paper, we study the correlation of the asymmetry and the
branching ratio, and then we will find only a type of interactions can be
enlarge the asymmetry. Therefore, in comparison with experiments, we have
possibility that we can constrain models beyond the standard model.
\end{abstract}

\section{Introduction}\label{sec:introduction}
The inclusive rare B decay $\bsll$ has already been studied by many
researchers. It is attractive to investigate this process experimentally or
theoretically. This decay mode is experimentally clean as well as $\bsg$,
specially in the low invariant mass region. And, when we can use a parton model
to study this process theoretically, because it is semileptonic decay. In the
standard model (SM), a flavor changing neutral current (FCNC) process appears
only through one or more loops. Since $\bsll$ is also a FCNC, new physics can
clarify itself to measure this decay. The extended models beyond the SM like the
minimal supersymmetrized model (MSSM) and the two Higgs doublets model (2HDM)
predict some deviation form the SM\cite{Goto} -\cite{Iltan}. The SM prediction
shows that, for $l=e$ or $\mu$, this mode will be found at the KEKB and the SLAC $e^+
e^-$ storage ring PEP-II B
factories in near future. Therefore, the study of this process is one of the
most interesting topic in order to search new physics. In this paper, the final
leptons will be a $\mu$ons or electrons throughly.

The CP-violating asymmetry of this decay is also a subject that many physicists
investigate. This observable is very sensitive to the complex phase of the CKM
matrix elements, so that we have the possibility to find effects beyond the SM. The
SM predicts that the CP asymmetry is suppressed, about $10^{-3}$ or
smaller\cite{DuYan,SMCPasym}. If some non-SM interactions enlarge for the
asymmetry to get sizable, we can know the existence beyond SM. This observable
has been calculated in MSSM and 2HDM\cite{LSCPasym} -\cite{Iltan}. In these
models, as well as the SM, the distribution is a function of fewer Wilson
coefficients than the full operator basis. In our previous work, we analyzed the
branching ratio and the forward-backward (FB) asymmetry, which is an observable
corresponding to the size of parity violation in the decay $\bsll$, with a most
general model-independent method\cite{FKMY,NONLocalFKY}. Generally, the matrix
element for the decay $b\to s l^+ l^-$ includes all types of local and
$bs\gamma$-induced four-Fermi operators. That is,
\bea
{\cal M} = \frac{G_F~ \alpha }{\sqrt{2}\pi }~V_{ts}^*V_{tb} 
              &[& C_{SL} ~\bar{s} i \sigma_{\mu \nu } \frac{q^\nu}{q^2} 
                        ( m_s L ) b 
                         ~\bar{l} \gamma^\mu l \nn \\
              &+& C_{BR} ~\bar{s} i \sigma_{\mu \nu } \frac{q^\nu}{q^2} 
                        ( m_b R ) b 
                        ~\bar{l} \gamma^\mu l  \nn \\
              &+& C_{LL} ~\bar{s}_L \gamma_\mu b_L 
                           ~\bar{l}_L \gamma^\mu l_L  \nn \\
              &+& C_{LR} ~\bar{s}_L \gamma_\mu b_L  
                           ~\bar{l}_R \gamma^\mu l_R  \nn \\
              &+& C_{RL} ~\bar{s}_R \gamma_\mu b_R 
                           ~\bar{l}_L \gamma^\mu l_L  \nn \\
              &+& C_{RR} ~\bar{s}_R \gamma_\mu b_R  
                           ~\bar{l}_R \gamma^\mu l_R  \nn \\
              &+& C_{LRLR} ~\bar{s}_L b_R ~\bar{l}_L l_R \nn \\
              &+& C_{RLLR} ~\bar{s}_R b_L ~\bar{l}_L l_R \nn \\
              &+& C_{LRRL} ~\bar{s}_L b_R ~\bar{l}_R l_L \nn \\  
              &+& C_{RLRL} ~\bar{s}_R b_L ~\bar{l}_R l_L \nn \\
              &+& C_T       ~\bar{s} \sigma_{\mu \nu } b 
                           ~\bar{l} \sigma^{\mu \nu } l \nn \\
              &+& i C_{TE}   ~\bar{s} \sigma_{\mu \nu } b 
                           ~\bar{l} \sigma_{\alpha \beta } l 
                           ~\epsilon^{\mu \nu \alpha \beta }],\label{eqn:matrix}
\eea
where $C_{XX}$'s are the coefficients of the four-Fermi interactions. Among
them, there are two $bs\gamma$ induced four-Fermi interactions denoted by
$C_{SL}$ and $C_{BR}$, which correspond to $-2 C_7^{eff}$ in the SM, and which
are constrained by the experimental data of $b\rightarrow s \gamma $. There are
four vector--type interactions denoted by $C_{LL}$, $C_{LR}$,   $C_{RL}$, and
$C_{RR}$. Two of them ($C_{LL}$, $C_{LR}$) are already present in the SM as the
combinations of ($C_9-C_{10}$, $C_9+C_{10}$). Therefore, they are regarded as
the sum of the contribution from the SM and the new physics deviations 
$(C_{LL}^{\rm new}, C_{LR}^{\rm new} )$. The other vector interactions denoted by
$C_{RL}$ and $C_{RR}$ are obtained by interchanging  the chirality projections
$L \leftrightarrow R $. There are four scalar--type interactions, $C_{LRLR}$,
$C_{RLLR}$, $C_{RLLR}$ and  $C_{RLRL}$. The remaining two denoted by $C_T$ and
$C_{TE}$ correspond to tensor--type. The indices, $L$ and $R$, are chiral
projections, $L =\frac{1}{2}( 1 - \gamma_5 ) $ and 
$R =\frac{1}{2}( 1 + \gamma_5)$. 
Then, we can get the differential branching ratio of the FCNC process 
$b \to s l^+ l^-$,
\bea
\frac{d {\cal B}}{d s } = \frac{1}{2{m_b}^8}~{\cal B}_0 &{\rm Re}[& 
             S_1(s) ~\{ m_s^2 |C_{SL}|^2  + m_b^2 |C_{BR}|^2 \} \nn \\
           &+& S_2(s) ~\{ 2 m_b m_s C_{SL} C_{BR}^*\} \nn \\
           &+& S_3(s)~\{ 2 m_s^2 C_{SL} ( C_{LL}^* + C_{LR}^* )
                   + 2 m_b m_s C_{BR} ( C_{RL}^* + C_{RR}^*)  \}
                     \nn \\
           &+& S_4(s)~\{ 2 m_b^2 C_{BR} ( C_{LL}^* + C_{LR}^* )
                   + 2 m_b m_s C_{SL} ( C_{RL}^* + C_{RR}^* )  \}
                     \nn \\ 
             &+& M_2(s)~\{ { \left|C_{LL}\right|^2 + \left|C_{LR}\right|^2 
             + \left|C_{RL}\right|^2 + \left|C_{RR}\right|^2  } \} \nn \\
             &+& M_6(s)~\{ - 2 { ( C_{LL} C_{RL}^* 
                                      + C_{LR} C_{RR}^* )} \nn \\
             & & ~~~~~~ + {(C_{LRLR} C_{RLLR}^* 
                            + C_{LRRL} C_{RLRL}^* ) } \} \nn \\
             &+& M_8(s)~\{ { \left|C_{LRLR}\right|^2 
                                + \left|C_{RLLR}\right|^2 
                                + \left|C_{LRRL}\right|^2        
                                + \left|C_{RLRL}\right|^2  } \} \nn \\
             &+& M_9(s)~\{ 16 { \left|C_{T}\right|^2 +
                           64 
                                \left|C_{TE}\right|^2 
                                } \} 
				],\label{eqn:branching}
\eea
Here, we ignore terms including lepton mass $m_l$, because we take only massless
(anti-) lepton into consideration. A set of the kinematic functions $S_i(s)$ 
($i = 1, 2, 3, 4, 5, 6$) and $M_n(s)$ ($n = 2, 6, 8$) is shown in Appendix
\ref{app:kinfun}. The normalization factor ${\cal B}_0$ is given by
\beq
{\cal B}_0 \equiv {\cal B}_{sl} \frac{3 \alpha^2}{16 \pi^2}
\frac{|V_{ts}^* V_{tb}|^2}{|V_{cb}|^2} \frac{1}{f(\hat{m_c})\kappa(\hat{m_c})},
\label{eqn:normalization}
\eeq
where the other factors $f(\hat{m_c})$ and $\kappa(\hat{m_c})$ are the phase
space factor and the $O(\alpha_s)$ QCD correction factor\cite{Kim}. The factor
${\cal B}_{sl}$ denotes the branching ratio of the semileptonic decay, and we
set it to $10.4 \%$. We can also have the FB asymmetry from
Eq.(\ref{eqn:matrix}). Thus, by numerical analysis, we got useful information to
pin down new physics beyond the standard model. However, we set all the new
Wilson coefficients to real when we carry out the numerical analysis. This means
that we assume that there is no new cp-violating source in the decay
$\bsll$. The CP asymmetry is sensitive to the imaginary part of the
coefficients. Therefore, it is worth treating the CP asymmetry based on the our
previous analysis.

This paper is organized as follows. In Section \ref{sec:cpasymmetry}, we find
the way how to get the general CP asymmetry, study the correlation between
the asymmetry and the branching ratio to pin down the type of interactions and
give some discussions. We give summary in Section \ref{sec:summary}.

\section{General CP Asymmetry}\label{sec:cpasymmetry}
We assume semileptonic decay $b \to c l^- \bar{\nu}_l$ is an approximately
CP-conserving mode, in fact experiments shows they corresponds
with each other within about $10^{-2}$\cite{PDG}. And, the partonic
approximation predicts no CP-violating asymmetry in the standard model
(SM). That is, we can use the same normalization factor as
Eq.(\ref{eqn:normalization}) to express the branching ratio of $b \to s l^+ l^-$
and $\bar{b} \to \bar{s} l^+ l^-$. 
For a
general Wilson coefficient $C_{XX}$, we can define $B_{XX}$, $\lambda_{XX}$ and
$A_{XX}$ by
\beq
C_{XX} \equiv B_{XX} + \lambda_{XX} A_{XX},\label{eqn:wilsondef}
\eeq
where $\lambda_{XX}$ is CP violating phase and generally the both of $B_{XX}$
and $A_{XX}$ are complex. In the case of the SM, only the CKM matrix elements
give the CP violating weak phase and the strong phase appears through QCD
penguin correction. Conventionally, these effects are included in the Wilson
coefficients $C_9^{eff}$ of the vector-type current-current
interaction\cite{Kruger}. Explicitly it is expressed by\cite{Kruger,Misiak}
\begin{equation}
C_9^{eff} = B_9 + \lambda_u A_9,\label{eqn:cqeff}
\end{equation}
where, without $c\bar{c}$ long-distant contribution,
\begin{equation}
B_9 = \left(1 + \alpha\frac{w(s)}{\pi}\right) C_9^{NDR} +
Y(s).\label{eqn:b9wilson}
\end{equation}
Only $\lambda_u \equiv (V_{ub} V{us}^*) / (V_{tb} V{ts}^*)$ includes
CP-violating phase. Since $\lambda_u A_9$ is very small except for $c \bar{c}$
resonance region, the SM predicts that the CP asymmetry is very
negligible\cite{DuYan}.

We must take $c \bar{c}$ resonance into consideration to discuss the branching
ratio and the CP asymmetry\cite{Long}, otherwise avoid region where $J / \psi$
and $\psi^{\prime}$ poles give contribution\cite{SMCPasym}. In this paper, we
take the latter stand. The residual region is lower region before $J / \psi$
resonance or higher region after $\psi^{\prime}$ resonance\cite{AvoidLong}. We
restrict our discussion to only low invariant mass region, $1 < s < 8$
(GeV$^2$), where $s \equiv (p_{l^+} + p_{l^-})^2$.  We then introduce the
partially integrated CP asymmetry ${\cal A}_{CP}$ defined by
\begin{equation}
{\cal A}_{CP} \equiv \frac{{\cal B}(\bsll) - {\cal B}(\absll)}{{\cal B}(\bsll) +
{\cal B}(\absll)} \equiv \frac{{\cal N}_{CP}}{{\cal
D}_{CP}},\label{eqn:partialCPasymm}
\end{equation}
where ${\cal B}(\bsll)$ is the partially integrated branching ratio for process
$\bsll$, defined by
\[\int^8_{1 (GeV^2)} ds \frac{d {\cal B}(\bsll)}{ds} \sim 3.73. \times 10^{-6}
~~(\mbox{at~} \mu = (m_b)_{\overline{MS}}).\]
In the same way, we define the partially integrated branching ratio for
$\absll$. We set $(C_7^{eff}, C_9^{NDR}, C_{10}) = (-0.317, 4.52, - 4.29)$ for numerical
calculation. We listed its value for the SM at renormalization scale 
$\mu = (m_b)_{\overline{MS}} = 4.2$ GeV in Table \ref{tab:partialcpasymmetrySM}, where
we set the Wolfenstein's CKM parameters\cite{Wolfen} to 
$(\rho, \eta) = (0.12, 0.25)$, $(0.16, 0.33)$ and $(0.27, 0.40)$. 
\begin{table}[h]
\begin{center}
\begin{tabular}{|c|c|}
\hline
$(\rho, \eta)$ & ${\cal A}_{CP}^{SM}$\\
\hline
$(0.12, 0.25)$ & $0.85 \times 10^{-3}$\\
$(0.16, 0.33)$ & $1.12 \times 10^{-3}$\\
$(0.27, 0.40)$ & $1.36 \times 10^{-3}$\\
\hline
\end{tabular}
\end{center}
\caption{The partially integrated CP asymmetry for $(\rho, \eta) = (0.12,
0.25)$, $(0.16, 0.33)$ and $(0.27, 0.40)$ and in the
SM at $\mu = (m_b)_{\overline{MS}}$.}\label{tab:partialcpasymmetrySM}
\end{table}
We should note that there is the huge uncertainty about the CP Asymmetry
predicted by the SM before we discuss the sensitivity to new
physics from our numerical results. The asymmetry in the SM is uncertain by
almost 100 $\%$ \cite{SMCPasym}. So, we must get at least $10$ times large size
as the SM prediction about the CP asymmetry to find the signal of new physics,
otherwise we fail to do.
Then, from Eq.(\ref{eqn:matrix}), we can get the
numerator ${\cal N}_{CP}$ of the CP asymmetry by replacing $Re(C_{XX} C_{YY}^*)$
in the branching ratio given in Eq.(\ref{eqn:branching}) with
\[- 2 Im(\lambda_{XX}) Im(B_{YY}^* A_{XX}) - 2 Im(\lambda_{YY}) Im(B_{XX}^*
A_{YY}) - 2 Im(\lambda_{XX} \lambda_{YY}^*) Im(A_{XX} A_{YY}^*),\]
and, for the dominator ${\cal D}_{CP}$, with
\[2 Re(B_{XX} B_{YY}^*) + 2 Re(\lambda_{XX}) Re(B_{YY}^* A_{XX}) + 2
Re(\lambda_{YY}) Re(B_{XX}^* A_{YY}) + 2 Re(\lambda_{XX} \lambda_{YY}^*) Re(A_{XX} A_{YY}^*).\]
The resultant CP asymmetry takes the most general model-independent
form. We show the explicit expression of this asymmetry in Appendix
\ref{app:kinfun}. The CP asymmetry does not vanish if and
only if $B_{XX}$ or $A_{XX}$ has deferent phase from $A_{YY}$ and $\lambda_{YY}$
or $\lambda_{XX} \lambda_{YY}^*$ has a imaginary part. Here, $XX$ and $YY$
denote types of interactions, whether they are the same type or not. 
However, in the most interesting models like the two higgs doublet model (2HDM)
\cite{AvoidLong,Iltan} and the minimal supersymmetrized standard model
(MSSM)\cite{Goto,Scimemi}, the strong phase does not play a so important role on
the CP asymmetry. Therefore, we assume that we can ignore a set of strong phase
introduced by new physics\cite{generalCPasym}. Then, for new vector, scalar and
tensor-type interactions, we can redefine the Wilson coefficients as
\begin{eqnarray}
C_{XX} &=& B_{XX}^{SM} + (\lambda_{XX} + \lambda_u) (A_9 + A_{XX})
\mbox{~~~for }XX = LL \mbox{ or } LR,\label{eqn:newlllrWilson}\\
C_{XX} &=& \lambda_{XX} A_{XX}
\mbox{~~~~~~~~~~~~~for others},\label{eqn:newlocalWilson}
\end{eqnarray}
Here, $A_{XX}$s are real and $\lambda_{XX}$s are phase factors defined by
$\exp(i \phi_{XX})$ where $0 \le \phi_{XX} < 2 \pi$, and 
$B_{LL}^{SM} \equiv B_9 - C_{10}$ and $B_{LR}^{SM} \equiv B_9 + C_{10}$. In the same
way, we can redefine $C_{BR}$ and $C_{SL}$, and have other constraints from the
measurement of $\bsg$,
\begin{equation}
	4 |C_7^{eff}|^2 (m_b^2 + m_s^2) = m_b^2 (|A_{SL}^N|^2 +
	|A_{BR}|^2)\label{nonlocalconstraint},
\end{equation}
where $A_{SL}^N = (m_b / m_s) A_{SL}$\cite{NONLocalFKY}. The definitions of $A_{BR}$
and $A_{SL}$, and $\phi_{BR}$ and $\phi_{SL}$, follows
Eq.(\ref{eqn:newlocalWilson}). Thus if there is the interference between such
coefficients and the $C_9^{eff}$, it can enlarge the CP asymmetry. Otherwise,
the new interactions suppress the observable under the above assumption. In this
case, the explicit form of the partially integrated CP asymmetry is given by
\beq
	{\cal A}_{CP} 
	\equiv \frac{\int^8_{1 GeV^2} ds (d{\cal N}_{CP}(s)/ds)}
		{\int^8_{1 GeV^2} ds (d{\cal D}_{CP}(s)/ds)}
	\equiv \frac{{\cal N}_{CP}}{{\cal D}_{CP}},
\eeq
where
\bea
\frac{d {\cal N}_{CP}(s)}{d s } &=&	- \frac{1}{{m_b}^8}~{\cal B}_0 [\nn\\ 
           &&\!\!\!\!\!\!\!\!\!\!\!\!\!\!\!\!\!\!\!\!\!\!\!\!\!\!\!\!\!\!\!\!\!\!\!\!\!
			S_3(s)~\{ 2 m_s^2 \left(Im(\lambda_{SL}) Im(A_{SL} B_9^*)
				+ Im(\lambda_{SL}\lambda_{LL}^*) Im(A_{SL} A_9^*)\right.\nn\\
			&&\!\!\!\!\!\!\!\!\!\!\!\!
			\left.+ Im(\lambda_{SL}) Im(A_{SL} B_9^*)
				+ Im(\lambda_{SL}\lambda_{LR}^*) Im(A_{SL} A_9^*)\right)\}\nn \\
           &&\!\!\!\!\!\!\!\!\!\!\!\!\!\!\!\!\!\!\!\!\!\!\!\!\!\!\!\!\!\!\!\!\!\!\!\!\!
			+ S_4(s)~\{ 2 m_b^2 \left(Im(\lambda_{BR}) Im(A_{BR} B_9^*)
				+ Im(\lambda_{BR}\lambda_{LL}^*) Im(A_{BR} A_9^*)\right.\nn\\
			&&\!\!\!\!\!
			\left. + Im(\lambda_{BR}) Im(A_{BR} B_9^*)
				+ Im(\lambda_{BR}\lambda_{LR}^*) Im(A_{BR} A_9^*)\right)\} \nn \\ 
             &&\!\!\!\!\!\!\!\!\!\!\!\!\!\!\!\!\!\!\!\!\!\!\!\!\!\!\!\!\!\!\!\!\!\!\!\!\!
			+ M_2(s)~\{ 2 \left( Im(\lambda_{LL}) Im((B_9 - C_{10}) (A_9 + A_{LL})^*)
				+ Im(\lambda_{LR}) Im((B_9 + C_{10}) (A_9 + A_{LR})^*)\right)\}\nn\\
             &&\!\!\!\!\!\!\!\!\!\!\!\!\!\!\!\!\!\!\!\!\!\!\!\!\!\!\!\!\!\!\!\!\!\!\!\!\!
			+ M_6(s)~\{ - 2 \left.( Im(\lambda_{RL}) Im((B_9^* - C_{10})A_{RL})
				+ Im(\lambda_{LL}\lambda_{RL}^*) Im((A_9 + A_{LL}) A_{RL}^*)\right.\nn\\
			&&\!\!\!\!\! 
				+\left.(Im(\lambda_{RR}) Im((B_9^* +C_{10}) A_{RR}) 
				+ Im(\lambda_{LR}\lambda_{RR}^*) Im((A_9 + A_{LR}) A_{RR}^*)\right)\}
			],\label{eqn:cpnumerator}
\eea
and
\bea
\frac{d {\cal D}_{CP}(s)}{d s } &=&	
			\frac{1}{{m_b}^8}~{\cal B}_0 [\nn\\ 
			&&\!\!\!\!\!\!\!\!\!\!\!\!\!\!\!\!\!\!\!\!\!\!\!\!\!\!\!\!\!\!\!\!
             S_1(s) ~\{ m_s^2 \left|A_{SL}\right|^2
				+  m_b^2 \left|A_{BR}\right|^2\} \nn \\
			&&\!\!\!\!\!\!\!\!\!\!\!\!\!\!\!\!\!\!\!\!\!\!\!\!\!\!\!\!\!\!\!\!
			+ S_2(s) ~\{ 2 m_b m_s Re(\lambda_{SL}\lambda_{BR}^*) Re(A_{SL} A_{BR}^*)\} \nn \\
           &&\!\!\!\!\!\!\!\!\!\!\!\!\!\!\!\!\!\!\!\!\!\!\!\!\!\!\!\!\!\!\!\!\!\!\!\!\!
			+ S_3(s)~\{ 2 m_s^2 \left(Re(\lambda_{SL}) Re(A_{SL} (B_9 - C_{10})^*)
				+ Re(\lambda_{SL}\lambda_{LL}^*) Re(A_{SL} (A_9 + A_{LL})^*)\right.\nn\\
			&&\!\!\!\!\!
			\left.+ Re(\lambda_{SL}) Re(A_{SL} (B_9 + C_{10})^*)
				+ Re(\lambda_{SL}\lambda_{LR}^*) Re(A_{SL} (A_9 + A_{LR})^*)\right)\nn\\
           &&\!\!\!\!\!\!\!\!\!\!\!\!\!\!\!\!\!\!
                   + 2 m_b m_s \left(Re(\lambda_{BR}\lambda_{RL}^*) Re(A_{BR}
				A_{RL}^*) + Re(\lambda_{BR}\lambda_{RR}^*) Re(A_{BR} A_{RR}^*)\right)\}\nn \\
           &&\!\!\!\!\!\!\!\!\!\!\!\!\!\!\!\!\!\!\!\!\!\!\!\!\!\!\!\!\!\!\!\!\!\!\!\!\!
			+ S_4(s)~\{ 2 m_b^2 \left(Re(\lambda_{BR}) Re(A_{BR} (B_9 - C_{10})^*)
				+ Re(\lambda_{BR}\lambda_{LL}^*) Re(A_{BR} (A_9 + A_{LL})^*)\right.\nn\\
			&&\!\!\!\!\!
			\left. + Re(\lambda_{BR}) Re(A_{BR} (B_9 + C_{10})^*)
				+ Re(\lambda_{BR}\lambda_{LR}^*) Re(A_{BR} (A_9 + A_{LR})^*)\right)\nn\\
           &&\!\!\!\!\!\!\!\!\!\!\!\!\!\!\!\!\!\!
                   + 2 m_b m_s \left(Re(\lambda_{SL}\lambda_{RL}^*) Re(A_{SL}
					A_{RL}^*) 
			+ Re(\lambda_{SL}\lambda_{RR}^*) Re(A_{SL} A_{RR}^*)\right)\} \nn \\ 
             &&\!\!\!\!\!\!\!\!\!\!\!\!\!\!\!\!\!\!\!\!\!\!\!\!\!\!\!\!\!\!\!\!\!\!\!\!\!
			+ M_2(s)~\{ \left|B_9 - C_{10}\right|^2 + \left|A_9 + A_{LL}\right|^2 
				+ 2 Re(\lambda_{LL}) Re((B_9 - C_{10}) (A_9 + A_{LL})^*)\nn\\
				&&\!\!\!\!\!\!\!\!\!\!\!\!\!\!\!
			+ \left|B_9 + C_{10}\right| + \left|A_9 + A_{LR}\right|^2
				+ 2 Re(\lambda_{LR}) Re((B_9 + C_{10}) (A_9 + A_{LR})^*)\nn\\
				&&\!\!\!\!\!\!\!\!\!\!\!\!\!\!\!
             + \left|A_{RL}\right|^2 + \left|A_{RR}\right|^2\} \nn \\
             &&\!\!\!\!\!\!\!\!\!\!\!\!\!\!\!\!\!\!\!\!\!\!\!\!\!\!\!\!\!\!\!\!\!\!\!\!\!
			+ M_6(s)~\{ - 2 \left.(Re(\lambda_{RL}) Re((B_9 - C_{10})^* A_{RL}) 
				+ Re(\lambda_{LL}\lambda_{RL}^*) Re((A_9 + A_{LL}) A_{RL}^*)\right.\nn\\
			&&\!\!\!\!\! 
				+\left.Re(\lambda_{RR}) Re((B_9 + C_{10})^* A_{RR}) 
				+ Re(\lambda_{LR}\lambda_{RR}^*) Re((A_9 + A_{LR}) A_{RR}^*)\right) \nn \\
           &&\!\!\!\!\!\!\!\!\!\!\!\!\!\!\!
			+ \left(Re(\lambda_{LRLR}\lambda_{RLLR}^*) Re(A_{LRLR} A_{RLLR}^*)\right.\nn\\
           &&\!\!\!\!\!\!\!\!\!\!\!\!\!\!\! \left.
              +Re(\lambda_{LRRL}\lambda_{RLRL}^*) Re(A_{LRRL} A_{RLRL}^*)\right)\}
                                \nn \\
             &&\!\!\!\!\!\!\!\!\!\!\!\!\!\!\!\!\!\!\!\!\!\!\!\!\!\!\!\!\!\!\!\!\!\!\!\!\!
				+M_8(s)~\{ \left|A_{LRLR}\right|^2
			+ \left|A_{RLLR}\right|^2 + \left|A_{LRRL}\right|^2 
			+ \left|A_{RLRL}\right|^2\} \nn \\
             &&\!\!\!\!\!\!\!\!\!\!\!\!\!\!\!\!\!\!\!\!\!\!\!\!\!\!\!\!\!\!\!\!\!\!\!\!\!
				+ M_9(s)~\{ 16 \left|A_T\right|^2 + 64 \left|A_{TE}\right|^2\}
				].\label{eqn:cpdenomitor}
\eea
Here, we omitted $\lambda_u$ because it is very small.

We will analyze the partially integrated CP asymmetry defined by
Eq.(\ref{eqn:partialCPasymm}) and examine its sensitivity to each Wilson
coefficient. For numerical estimation, we set $(\rho, \eta) = (0.16, 0.33)$. At
first, we investigate vector, scalar and tensor-type interactions, which are
collectively {\it new local interactions}. The results of Ref. \cite{FKMY} make
us predict the sensitivity of the CP asymmetry to each Wilson coefficient. The
branching ratio is the most sensitive to the vector-type interactions, specially
$C_{LL}$, and the contribution due to $C_{RL}$ and $C_{RR}$ it positive. And
only the $C_{LL}$ and $C_{LR}$ have the weak and the strong phase, so we can
expect that only the two types of interactions can make CP asymmetry be large,
specially we can expect that the CP asymmetry is sizable by appropriate
$C_{LL}$. However, $C_{RL}$ and $C_{RR}$ would suppress the CP asymmetry. The
scalar and tensor-type interactions hardly interfere with each other or a
vector-type interaction in the massless lepton limit. Thus, if a scalar or
tensor-type interaction enters into our decay mode, it would suppress the CP
asymmetry. In Figures \ref{fig:correlationLL} - \ref{fig:correlationLR}, I
plotted the correlation between the branching ratio and the CP asymmetry when
$C_{LL}$ or $C_{LR}$ moves. Because the flow of each interaction depends on
the type of the interaction, we can pin down the type of interaction which
contributes to the processes once we measured these observable. These show
behavior as expected in the above discussion. We should take attention to Figure
\ref{fig:correlationLL}, which shows the CP asymmetry can get much larger as the
branching ratio is about predicted by the SM. It is because the partially
integrated CP asymmetry for the SM is so suppressed why it is enlarged by
$10^2$. Extremely, for $\phi_{LL} = \pi / 4$, $\pi / 2$ or $3 \pi / 4$, the
asymmetry is the most enlarged when $A_{LL} \sim - 1.2 |C_{10}|$, $0$ or 
$1.1 |C_{10}|$. If we ignore the SM CP-violating contribution, $A_9$ and 
$\lambda_u$, $C_{LL}$ enters into the asymmetry as followings
\begin{equation}
\frac{2 \int ds (Im(M_2 (B_9 - C_{10}) - 2 M_4 C_7^{eff}) (A_9^* + A_{LL}))
	\sin\phi_{LL}}
{2 m_b^8 {\cal B}_{SM} + \int ds M_2 |A_9 + A_{LL}|^2 
	+ 2 \int ds Re(M_2 (B_9 - C_{10}) - 2 M_4 C_7^{eff}) (A_9^* + A_{LL}))
	\cos\phi_{LL}},
\label{eqn:cpasymmetryfromCLL}
\end{equation}
where $M_2(s)$ and $M_4(s)$ are shown in Appendix \ref{app:kinfun}, and 
$2 m_b^8 {\cal B}_{SM} \sim 0.72$. By choosing an approximate set of $A_{LL}$
and $\phi_{LL}$ to hold
\[\int ds M_2 A_{LL} \sim - 2 \int ds M_2 Re(B_9 - C_{10})
\cos\phi_{LL},\]
the asymmetry can get $10^{-1}$. That is, if there is new physics through
$C_{LL}$ with weak phase, we have the possibility that we may pin down this type
of interactions at the B factory in near future even if there is no contradiction
with present experiments. And, Eq.(\ref{eqn:cpasymmetryfromCLL}) shows the
correlation is very sensitive to whether $\phi_{LL}$ is infinitesimal or
not. Thus the SM prediction point is far from other lines. In the same way, some
$A_{LR}$ and $\phi_{LR}$ enlarge the asymmetry and sensitive to $\phi_{LR}$ but,
because $B_9 + C_{10} \ll B_9 - C_{10}$, its contribution is smaller than
$A_{LL}$ and $\phi_{LL}$. And, in order to see how much the coefficient $A_{LR}$
contribute to the asymmetry, we check at which the absolute value of the
asymmetry becomes the maximum. By analogy with the analysis for $A_{LL}$ and
Eq.(\ref{eqn:cpasymmetryfromCLL}), we find that it has the largest value when,
roughly,
\[ 2 \int ds M_2 Re(A_9) \sim - \int ds M_2 A_{LR}, \]
numerically $A_{LR} \sim - 1.4 |C_{10}|$ or $- 1.5 |C_{10}|$ for 
$\phi_{LR} = \pi / 4$ or $\pi / 2$ and $3 \pi / 4$. (Note we ignored the term
including $M_2 Re(B_9 + C_{10}) - 2 M_4 C_7^{eff}$ because it is much smaller
than the remain in the dominator.) 

For $C_{RL}$ and $C_{RR}$, the terms from $M_2 C_{RL}^2$ and $M_2 C_{RR}^2$
disappear in the numerator, so that the other terms,
\begin{eqnarray}
&&-M_6(s) (C_9^{eff} - C_{10}) C_{RL}^*,\label{eqn:crlsense}\\
&&-M_6(s) (C_9^{eff} + C_{10}) C_{RL}^*,\label{eqn:crrsense}
\end{eqnarray}
which we ignored when we had discussed the sensitivity of $C_{RL}$ and $C_{RR}$
to the branching ratio, give significant effect to the CP asymmetry, so that the
asymmetry may depend on $\phi_{RL}$ and $\phi_{RR}$. Here, $M_6(s)$ is given in
Appendix \ref{app:kinfun}. Eqs.(\ref{eqn:crlsense}) and (\ref{eqn:crrsense}) give similar
contribution to the asymmetry except that it includes not $M_2$ but $M_6$. Since
$M_6 \ll M_2$ due to strange quark mass $m_s$, its sensitivity is tiny. 
We can also consider the correlation where $A_{RL}$ and $A_{RR}$ are
very small strong phase, that is, 
\begin{eqnarray}
	A_{RL} &=& A_9 + A_{RL}^{\prime},\\
	A_{RR} &=& A_9 + A_{RR}^{\prime},\\
\end{eqnarray}
where $A_{RL}^{\prime}$ and $A_{RR}^{\prime}$ are real. In this case, the sign
of the imaginary part of the $(B_9 - C_{10}) (A_9 + A_{RL}^{\prime})$ and 
$(B_9 + C_{10}) (A_9 + A_{RR}^{\prime})$ yields the difference between the
correlations, however, the sensitivity is still tiny.

For scalar and tensor interactions, in the massless lepton limit, their Wilson
coefficients appear only through the squared absolute. So, the asymmetry is
almost independent of $\phi_S$ ($S = LRLR$, $LRRL$, $RLLR$, $RLRL$), $\phi_T$ and
$\phi_{TE}$ and it gets only more suppressed as $A_S$, $A_T$ or $A_{TE}$ gets
larger. Moreover, the sensitivity is very small because the corresponding
kinematic functions include a factor $m_l$.
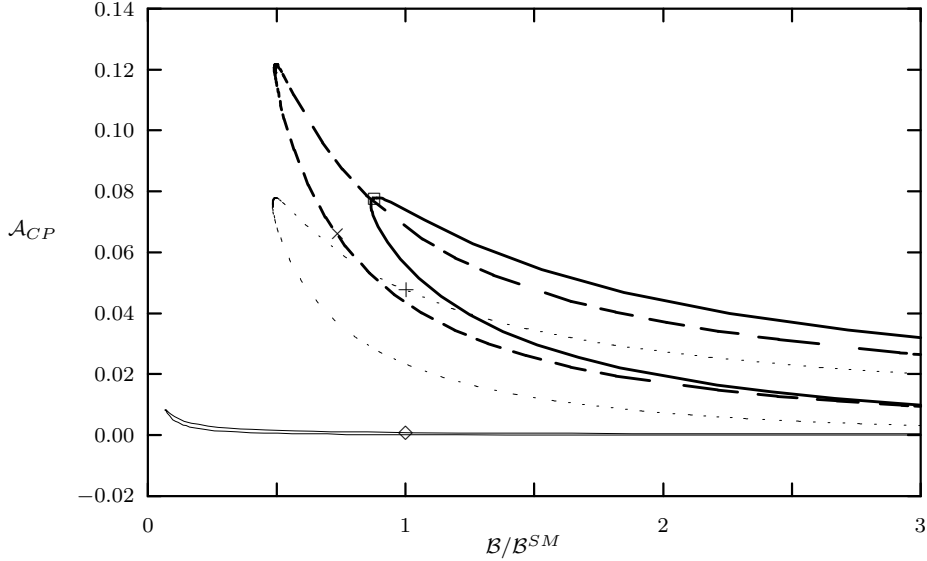
\begin{figure}
\setlength{\unitlength}{0.240900pt}
\begin{picture}(1500,900)(0,0)
\footnotesize
\thicklines \path(220,90)(240,90)
\thicklines \path(1433,90)(1413,90)
\put(198,90){\makebox(0,0)[r]{$-0.02$}}
\thicklines \path(220,186)(240,186)
\thicklines \path(1433,186)(1413,186)
\put(198,186){\makebox(0,0)[r]{$0.00$}}
\thicklines \path(220,282)(240,282)
\thicklines \path(1433,282)(1413,282)
\put(198,282){\makebox(0,0)[r]{$0.02$}}
\thicklines \path(220,377)(240,377)
\thicklines \path(1433,377)(1413,377)
\put(198,377){\makebox(0,0)[r]{$0.04$}}
\thicklines \path(220,473)(240,473)
\thicklines \path(1433,473)(1413,473)
\put(198,473){\makebox(0,0)[r]{$0.06$}}
\thicklines \path(220,569)(240,569)
\thicklines \path(1433,569)(1413,569)
\put(198,569){\makebox(0,0)[r]{$0.08$}}
\thicklines \path(220,665)(240,665)
\thicklines \path(1433,665)(1413,665)
\put(198,665){\makebox(0,0)[r]{$0.10$}}
\thicklines \path(220,760)(240,760)
\thicklines \path(1433,760)(1413,760)
\put(198,760){\makebox(0,0)[r]{$0.12$}}
\thicklines \path(220,856)(240,856)
\thicklines \path(1433,856)(1413,856)
\put(198,856){\makebox(0,0)[r]{$0.14$}}
\thicklines \path(220,90)(220,110)
\thicklines \path(220,856)(220,836)
\put(220,45){\makebox(0,0){$0$}}
\thicklines \path(422,90)(422,110)
\thicklines \path(422,856)(422,836)
\thicklines \path(624,90)(624,110)
\thicklines \path(624,856)(624,836)
\put(624,45){\makebox(0,0){$1$}}
\thicklines \path(826,90)(826,110)
\thicklines \path(826,856)(826,836)
\thicklines \path(1029,90)(1029,110)
\thicklines \path(1029,856)(1029,836)
\put(1029,45){\makebox(0,0){$2$}}
\thicklines \path(1231,90)(1231,110)
\thicklines \path(1231,856)(1231,836)
\thicklines \path(1433,90)(1433,110)
\thicklines \path(1433,856)(1433,836)
\put(1433,45){\makebox(0,0){$3$}}
\thicklines \path(220,90)(1433,90)(1433,856)(220,856)(220,90)
\thinlines \path(1433,186)(1391,186)(1265,186)(1153,186)(1048,186)(957,186)(871,186)(782,186)(704,187)(663,187)(627,187)(564,187)(531,187)(501,188)(472,188)(446,188)(424,189)(402,189)(381,190)(360,191)(341,192)(324,193)(308,195)(295,197)(285,199)(280,201)(275,202)(267,206)(263,208)(259,210)(254,216)(250,221)(249,223)(248,224)(248,225)(248,225)(247,225)(247,226)(247,226)(247,226)(247,226)(247,226)(247,226)(248,226)(248,226)(248,226)(248,226)(248,226)(248,226)(249,225)(251,223)
\thinlines \path(251,223)(256,219)(262,215)(268,212)(277,208)(288,205)(299,203)(314,200)(330,198)(346,197)(365,196)(383,195)(404,194)(451,193)(479,192)(507,191)(565,191)(635,190)(708,189)(792,189)(885,189)(980,188)(1179,188)(1291,188)(1417,188)(1433,188)
\thinlines \dashline[-10]{5}(1433,201)(1337,204)(1241,208)(1158,212)(1071,218)(995,224)(857,240)(796,250)(735,263)(682,277)(630,295)(584,316)(545,340)(511,368)(485,396)(460,429)(442,461)(429,493)(423,508)(419,523)(418,529)(416,535)(416,540)(415,542)(415,544)(415,545)(415,546)(415,547)(415,548)(415,548)(415,549)(415,549)(415,550)(415,550)(415,551)(415,552)(415,552)(415,553)(415,554)(415,554)(416,555)(416,556)(416,556)(416,557)(417,557)(417,558)(417,558)(417,558)(418,558)(418,558)(418,558)
\thinlines \dashline[-10]{5}(418,558)(418,559)(419,559)(419,559)(419,559)(420,559)(420,559)(420,559)(421,559)(421,558)(422,558)(423,558)(425,557)(427,555)(433,550)(469,515)(529,466)(566,442)(606,421)(699,384)(757,367)(814,353)(883,339)(951,327)(1103,308)(1277,292)(1433,282)
\Thicklines \path(1433,233)(1393,236)(1298,244)(1198,254)(1113,265)(961,292)(896,308)(831,328)(777,349)(725,375)(680,404)(645,433)(616,463)(596,489)(581,514)(575,527)(573,532)(572,537)(571,539)(571,541)(570,543)(570,545)(570,546)(570,546)(570,547)(570,548)(570,548)(570,549)(570,549)(570,550)(570,551)(570,552)(570,552)(570,553)(570,554)(571,555)(571,556)(572,557)(572,557)(573,558)(574,558)(574,559)(575,559)(575,559)(576,559)(576,559)(577,559)(577,559)(578,559)(579,559)(579,559)
\Thicklines \path(579,559)(580,559)(580,559)(581,559)(581,559)(582,559)(583,559)(585,559)(588,558)(592,556)(601,552)(654,524)(730,487)(838,446)(967,410)(1133,377)(1319,351)(1433,339)
\Thicklines \dashline[-20]{1}(1433,231)(1403,233)(1301,240)(1208,247)(1114,257)(954,279)(883,293)(813,311)(753,330)(701,351)(649,378)(606,407)(567,441)(529,483)(498,528)(471,579)(449,634)(433,683)(428,704)(424,725)(421,740)(420,746)(419,752)(419,755)(419,756)(418,757)(418,758)(418,759)(418,760)(418,761)(418,762)(418,763)(418,763)(418,764)(418,764)(419,765)(419,765)(419,766)(419,766)(419,767)(419,767)(419,768)(419,768)(420,768)(420,768)(420,768)(420,768)(421,768)(421,768)(421,768)(422,768)
\Thicklines \dashline[-20]{1}(422,768)(422,768)(423,766)(425,763)(428,760)(433,750)(449,720)(494,645)(525,604)(563,562)(651,494)(702,464)(762,437)(884,396)(954,379)(1034,363)(1115,349)(1210,336)(1400,315)(1433,312)
\put(624,190){\raisebox{-1.2pt}{\makebox(0,0){$\Diamond$}}}
\put(625,414){\makebox(0,0){$+$}}
\put(575,559){\raisebox{-1.2pt}{\makebox(0,0){$\Box$}}}
\put(517,502){\makebox(0,0){$\times$}}
\put(0,500){\mbox{${\cal A}_{CP}$}}
\put(750,0){\mbox{${\cal B} / {\cal B}^{SM}$}}
\end{picture}
\caption{The correlation between ${\cal B} / {\cal B}^{SM}$ and
${\cal A}_{CP}$ as $A_{LL}$ moves, and $\phi_{LL} = 0$
(thin solid line), $\pi / 4$ (dotted line), $\pi / 2$ (thick solid line) and 
$3 \pi / 4$ (dashed line). The marks $\Diamond$, $+$, $\Box$ and $\times$ show
the prediction for $\phi_{LL} = 0$, $\pi / 4$, $\pi / 2$ and $3 \pi / 4$ with
$A_{LL} = 0$.}\label{fig:correlationLL}
\end{figure}
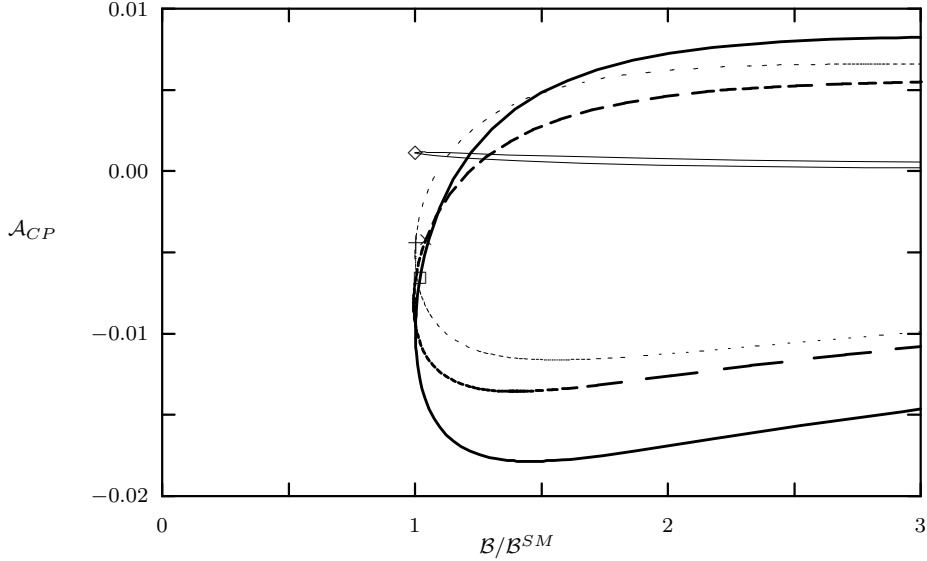
\begin{figure}
\setlength{\unitlength}{0.240900pt}
\begin{picture}(1500,900)(0,0)
\footnotesize
\thicklines \path(242,90)(262,90)
\thicklines \path(1433,90)(1413,90)
\put(220,90){\makebox(0,0)[r]{$-0.02$}}
\thicklines \path(242,218)(262,218)
\thicklines \path(1433,218)(1413,218)
\thicklines \path(242,345)(262,345)
\thicklines \path(1433,345)(1413,345)
\put(220,345){\makebox(0,0)[r]{$-0.01$}}
\thicklines \path(242,473)(262,473)
\thicklines \path(1433,473)(1413,473)
\thicklines \path(242,601)(262,601)
\thicklines \path(1433,601)(1413,601)
\put(220,601){\makebox(0,0)[r]{$0.00$}}
\thicklines \path(242,728)(262,728)
\thicklines \path(1433,728)(1413,728)
\thicklines \path(242,856)(262,856)
\thicklines \path(1433,856)(1413,856)
\put(220,856){\makebox(0,0)[r]{$0.01$}}
\thicklines \path(242,90)(242,110)
\thicklines \path(242,856)(242,836)
\put(242,45){\makebox(0,0){$0$}}
\thicklines \path(441,90)(441,110)
\thicklines \path(441,856)(441,836)
\thicklines \path(639,90)(639,110)
\thicklines \path(639,856)(639,836)
\put(639,45){\makebox(0,0){$1$}}
\thicklines \path(838,90)(838,110)
\thicklines \path(838,856)(838,836)
\thicklines \path(1036,90)(1036,110)
\thicklines \path(1036,856)(1036,836)
\put(1036,45){\makebox(0,0){$2$}}
\thicklines \path(1235,90)(1235,110)
\thicklines \path(1235,856)(1235,836)
\thicklines \path(1433,90)(1433,110)
\thicklines \path(1433,856)(1433,836)
\put(1433,45){\makebox(0,0){$3$}}
\thicklines \path(242,90)(1433,90)(1433,856)(242,856)(242,90)
\thinlines \path(1433,606)(1347,606)(1174,608)(1097,609)(1021,610)(903,613)(851,615)(801,617)(760,619)(722,621)(671,625)(654,627)(648,628)(643,628)(642,629)(640,629)(640,629)(640,629)(639,629)(639,629)(639,630)(639,630)(639,630)(639,630)(639,630)(639,630)(639,630)(639,630)(639,630)(639,630)(639,630)(639,630)(640,630)(640,630)(640,630)(641,630)(641,630)(642,630)(643,630)(644,630)(645,630)(646,631)(646,631)(647,631)(648,631)(648,631)(649,631)(649,631)(650,631)(651,631)(651,631)
\thinlines \path(651,631)(652,631)(652,631)(653,631)(654,631)(655,631)(657,630)(661,630)(665,630)(685,630)(712,629)(785,626)(878,624)(1002,621)(1146,618)(1308,616)(1433,615)
\thinlines \dashline[-10]{5}(1433,348)(1371,343)(1176,326)(1089,318)(1001,311)(965,308)(945,307)(927,306)(911,305)(903,305)(894,305)(886,305)(879,304)(876,304)(872,304)(869,304)(866,304)(862,304)(858,304)(855,304)(851,304)(848,304)(844,304)(837,305)(831,305)(825,305)(812,306)(805,306)(798,307)(786,308)(774,310)(761,313)(750,315)(739,318)(721,325)(704,334)(687,346)(675,358)(664,373)(659,382)(654,392)(650,401)(647,410)(644,421)(642,431)(642,436)(641,442)(640,448)(640,452)(640,454)(640,457)
\thinlines \dashline[-10]{5}(640,457)(640,460)(639,463)(639,464)(639,466)(639,467)(639,469)(639,472)(639,473)(639,475)(639,476)(640,478)(640,481)(640,484)(640,489)(641,495)(641,500)(643,512)(645,525)(651,548)(659,569)(668,589)(692,627)(709,647)(727,664)(748,680)(769,694)(816,716)(843,725)(873,734)(934,747)(967,752)(1005,756)(1043,760)(1080,762)(1122,765)(1164,766)(1188,767)(1211,768)(1238,768)(1264,768)(1287,769)(1301,769)(1314,769)(1321,769)(1328,769)(1334,769)(1340,769)(1344,769)(1347,769)(1351,769)
\thinlines \dashline[-10]{5}(1351,769)(1355,769)(1362,769)(1366,769)(1369,769)(1375,769)(1381,769)(1393,769)(1406,769)(1419,769)(1433,769)
\Thicklines \path(1433,227)(1404,223)(1245,201)(1110,181)(985,161)(934,154)(883,148)(870,147)(857,146)(846,146)(840,145)(834,145)(832,145)(829,145)(826,145)(824,145)(821,145)(819,145)(817,145)(814,145)(812,145)(810,145)(809,145)(804,145)(802,145)(799,145)(794,146)(784,146)(775,148)(766,149)(756,151)(741,155)(725,161)(711,168)(700,176)(688,186)(678,199)(669,212)(661,227)(655,243)(650,259)(648,268)(646,278)(643,297)(642,308)(641,318)(640,324)(640,330)(640,333)(640,337)(640,338)
\Thicklines \path(640,338)(640,340)(640,343)(640,346)(640,348)(640,349)(640,351)(640,352)(640,354)(640,356)(640,359)(640,361)(640,368)(641,374)(641,381)(642,395)(643,407)(645,420)(649,445)(654,469)(661,494)(678,544)(699,588)(727,630)(760,667)(796,698)(837,724)(879,743)(925,760)(981,775)(1039,786)(1106,795)(1140,798)(1178,801)(1213,804)(1252,806)(1292,808)(1330,809)(1371,810)(1395,810)(1417,811)(1433,811)
\Thicklines \dashline[-20]{1}(1433,325)(1315,312)(1166,295)(1040,279)(925,264)(878,259)(855,257)(832,256)(821,255)(815,255)(812,255)(809,255)(807,255)(804,255)(802,255)(799,255)(798,255)(797,255)(794,255)(792,255)(791,255)(789,255)(787,255)(785,255)(781,255)(776,255)(771,255)(762,256)(753,257)(746,258)(738,259)(731,261)(723,263)(710,267)(699,272)(688,278)(677,286)(667,296)(659,306)(652,318)(647,331)(643,343)(641,350)(640,357)(639,364)(638,368)(638,372)(638,376)(637,380)(637,382)(637,384)
\Thicklines \dashline[-20]{1}(637,384)(637,385)(637,387)(637,388)(637,389)(637,391)(637,393)(637,394)(637,395)(637,396)(637,399)(637,400)(637,401)(637,405)(638,409)(638,414)(639,423)(640,431)(642,441)(645,458)(650,476)(657,495)(672,530)(694,565)(721,597)(752,624)(788,647)(824,666)(866,682)(917,697)(969,708)(1030,718)(1097,726)(1130,729)(1166,731)(1200,733)(1238,735)(1310,738)(1350,739)(1389,740)(1432,741)(1433,741)
\put(639,630){\raisebox{-1.2pt}{\makebox(0,0){$\Diamond$}}}
\put(640,489){\makebox(0,0){$+$}}
\put(647,434){\raisebox{-1.2pt}{\makebox(0,0){$\Box$}}}
\put(656,493){\makebox(0,0){$\times$}}
\put(0,500){\mbox{${\cal A}_{CP}$}}
\put(740,0){\mbox{${\cal B} / {\cal B}^{SM}$}}
\end{picture}
\caption{The correlation between ${\cal B} / {\cal B}^{SM}$ and
${\cal A}_{CP}$ as $A_{LR}$ moves, and $\phi_{LR} = 0$
(thin solid line), $\pi / 4$ (dotted line), $\pi / 2$ (thick solid line) and
$3 \pi / 4$ (dashed line). The mark $\Diamond$ shows the standard model
prediction. The marks $\Diamond$, $+$, $\Box$ and $\times$ show the
prediction for $\phi_{LR} = 0$, $\pi / 4$, $\pi / 2$ and $3 \pi / 4$ with 
$A_{LR} = 0$.}\label{fig:correlationLR}
\end{figure}

Next, consider only $C_{BR}$ and $C_{SL}$, which is constrained by
Eq.(\ref{nonlocalconstraint}). Generally, without strong phase, these
coefficients are expressed by
\begin{equation}
	C_{BR} = A_{BR} e^{i \phi_{BR}}, ~~~~
	C_{SL} = A_{SL} e^{i \phi_{SL}},\label{eqn:generalbrsl}
\end{equation}
where $\phi_{BR}$ and $\phi_{SL}$ are independent weak phases. As shown in
Ref. \cite{NONLocalFKY}, the partially integrated branching ratio ${\cal B}$ is
more sensitive to $C_{BR}$ than $C_{SL}^N \equiv (m_b / m_s) C_{SL}$ because of
the strange quark mass $m_s$. This is true for the partially integrated CP
asymmetry ${\cal A}_{CP}$. In other words, it is almost independent of the phase
$\phi_{SL}$ in comparison with $\phi_{BR}$. And the asymmetry cannot be so
enlarged by $A_{SL}$ (or $A_{BR}$) with $\phi_{BR} = 0$. We can find this
feature by comparing Figure \ref{fig:correlationBR} with Figure
\ref{fig:correlationNL}. In the former, we set $\phi_{SL}$ to 0, in the latter,
however, we set $\phi_{SL} = \phi_{BR} \equiv \phi_{NL}$ 
By contrast with $C_{SL}$, the form of the correlation depends on $C_{BR}$
considerably. Ignoring the SM contribution, in the case of 
$\phi_{SL} = \phi_{BR} = \phi_{NL}$, the asymmetry takes a form of 
\begin{equation}
\frac{8 m_b C_7^{eff} \left(m_s \int ds S_3 \cos\theta Im(B_9)
	+ m_b \int ds S_4 \sin\theta Im(B_9)\right) \sin\phi_{NL}}
	{2 m_b^8 {\cal B}_{NL} + 2 m_b^8 {\cal B}_{L} 
	- 8 m_b C_7^{eff} \left(m_s \int ds S_3 \cos\theta Re(B_9)
	+ m_b \int ds \sin\theta S_4 Re(B_9)\right)
	\cos\phi_{NL}},\label{eqn:cpasymmetryNLonly}
\end{equation}
where ${\cal B}_{NL}$ and ${\cal B}_L$ are the partially integrated branching
ratios. For the former, only non-vanishing new Wilson coefficients are
$A_{BR}$ and the latter has $A_{SL}$ and  $A_{BR} = A_{SL} = 0$. We set
$\tan\theta = A_{BR} / A_{SL}^N$ and ignored the higher order terms about 
$m_s / m_b$. The definition of $S_3$ and $S_4$ is given in Appendix
\ref{app:kinfun}. Since $Im(B_9) \ll Re(B_9)$, the partially integrated
branching ratio is expressed by 
\begin{equation}
\frac{1}{2 m_b^8} {\cal B}_0 \left[{\cal B}_{NL} + {\cal B}_{L}
	- 8 m_b C_7^{eff} \left(m_s \int ds S_3 Re(B_9) \cos\theta
	+ m_b \int ds S_4 Re(B_9) \sin\theta\right) \cos\phi_{NL}\right].
\label{eqn:nonlocalbratio}
\end{equation}
Eqs.(\ref{eqn:cpasymmetryNLonly}) and (\ref{eqn:nonlocalbratio}) show that, when
$\phi_{NL}$ rounds from $0$ to $2\pi$, the ellipse of the correlation, as shown
in Figure \ref{fig:correlationNL}, does. The size of $A_{BR}$, and also
$A_{SL}$, is not so significant to enlarge the partially integrated CP
asymmetry. Thus, these two types of interaction do not give so great influence
to the partially integrated CP asymmetry even if there is another type of new
interactions, say $C_{LL}$. For example, when we set $\phi_{LL} = \pi / 2$ and
we check the dependency of $\phi_{NL}$ on the asymmetry, it does not so largely
changes the form of the correlation between ${\cal B}$ and ${\cal A}$ as
$A_{LL}$ moves negligible, as shown in Figure \ref{fig:correlationLLNL}, so we
must note if and only if very minute experiments were done.
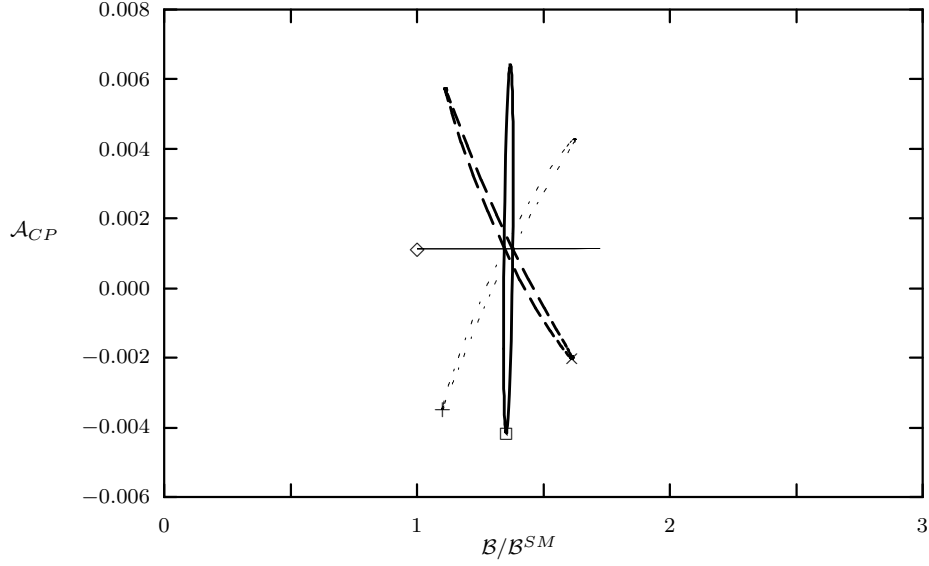
\begin{figure}
\setlength{\unitlength}{0.240900pt}
\begin{picture}(1500,900)(0,0)
\footnotesize
\thicklines \path(242,90)(262,90)
\thicklines \path(1433,90)(1413,90)
\put(220,90){\makebox(0,0)[r]{$-0.006$}}
\thicklines \path(242,199)(262,199)
\thicklines \path(1433,199)(1413,199)
\put(220,199){\makebox(0,0)[r]{$-0.004$}}
\thicklines \path(242,309)(262,309)
\thicklines \path(1433,309)(1413,309)
\put(220,309){\makebox(0,0)[r]{$-0.002$}}
\thicklines \path(242,418)(262,418)
\thicklines \path(1433,418)(1413,418)
\put(220,418){\makebox(0,0)[r]{$0.000$}}
\thicklines \path(242,528)(262,528)
\thicklines \path(1433,528)(1413,528)
\put(220,528){\makebox(0,0)[r]{$0.002$}}
\thicklines \path(242,637)(262,637)
\thicklines \path(1433,637)(1413,637)
\put(220,637){\makebox(0,0)[r]{$0.004$}}
\thicklines \path(242,747)(262,747)
\thicklines \path(1433,747)(1413,747)
\put(220,747){\makebox(0,0)[r]{$0.006$}}
\thicklines \path(242,856)(262,856)
\thicklines \path(1433,856)(1413,856)
\put(220,856){\makebox(0,0)[r]{$0.008$}}
\thicklines \path(242,90)(242,110)
\thicklines \path(242,856)(242,836)
\put(242,45){\makebox(0,0){$0$}}
\thicklines \path(441,90)(441,110)
\thicklines \path(441,856)(441,836)
\thicklines \path(639,90)(639,110)
\thicklines \path(639,856)(639,836)
\put(639,45){\makebox(0,0){$1$}}
\thicklines \path(838,90)(838,110)
\thicklines \path(838,856)(838,836)
\thicklines \path(1036,90)(1036,110)
\thicklines \path(1036,856)(1036,836)
\put(1036,45){\makebox(0,0){$2$}}
\thicklines \path(1235,90)(1235,110)
\thicklines \path(1235,856)(1235,836)
\thicklines \path(1433,90)(1433,110)
\thicklines \path(1433,856)(1433,836)
\put(1433,45){\makebox(0,0){$3$}}
\thicklines \path(242,90)(1433,90)(1433,856)(242,856)(242,90)
\thinlines \path(789,480)(789,480)(752,480)(715,480)(684,480)(661,480)(652,480)(645,480)(642,480)(640,480)(640,480)(639,480)(639,480)(639,480)(639,480)(639,480)(639,480)(639,480)(639,480)(639,480)(639,480)(639,480)(639,480)(639,480)(639,480)(639,480)(639,480)(639,480)(639,480)(639,480)(639,480)(639,480)(640,480)(640,480)(640,480)(640,480)(640,480)(640,480)(640,480)(640,480)(641,480)(641,480)(641,480)(641,480)(641,480)(642,480)(642,480)(643,480)(644,480)(646,480)(649,480)
\thinlines \path(649,480)(657,480)(679,480)(709,480)(742,480)(781,480)(818,481)(852,481)(868,481)(875,481)(883,481)(889,481)(895,481)(898,481)(899,481)(900,481)(902,481)(902,481)(903,481)(903,481)(904,481)(905,481)(905,481)(906,481)(906,481)(906,481)(907,481)(907,481)(908,481)(908,481)(910,481)(911,481)(912,481)(914,481)(915,481)(917,481)(920,481)(921,481)(922,481)(923,481)(924,481)(925,481)(925,481)(926,481)(926,481)(926,481)(926,481)(926,481)(926,481)(926,481)(926,481)
\thinlines \path(926,481)(926,481)(926,481)(926,481)(926,481)(926,481)(926,481)(926,481)(926,481)(926,481)(925,481)(924,481)(922,481)(917,481)(908,481)(887,480)(857,480)(823,480)(789,480)
\thinlines \dashline[-10]{5}(789,480)(789,480)(762,427)(735,367)(713,313)(695,270)(683,239)(680,231)(679,229)(679,229)(679,228)(679,228)(679,228)(679,228)(679,228)(678,228)(678,228)(678,228)(678,228)(678,228)(678,228)(678,228)(678,228)(678,228)(678,228)(678,228)(678,228)(678,228)(678,229)(678,229)(678,229)(678,229)(679,231)(681,238)(689,267)(705,312)(725,365)(750,425)(776,481)(802,530)(828,575)(850,609)(869,634)(876,643)(881,649)(884,652)(885,652)(886,652)(886,653)(886,653)(886,653)
\thinlines \dashline[-10]{5}(886,653)(886,653)(886,653)(886,653)(886,653)(886,653)(886,653)(886,653)(886,653)(886,653)(886,653)(887,652)(887,652)(887,652)(887,652)(887,652)(887,652)(887,652)(887,652)(887,652)(887,652)(886,651)(886,651)(886,649)(885,647)(877,633)(861,607)(840,572)(816,531)(789,480)
\Thicklines \path(789,480)(789,480)(789,445)(788,408)(787,335)(785,278)(784,251)(783,231)(782,214)(781,202)(780,197)(780,195)(780,194)(780,193)(780,192)(779,192)(779,191)(779,191)(779,191)(779,191)(779,191)(779,191)(779,191)(779,191)(779,191)(779,191)(779,191)(779,191)(779,191)(779,191)(779,191)(779,191)(778,192)(778,193)(778,194)(778,195)(778,199)(777,204)(777,210)(776,219)(776,228)(776,238)(776,249)(775,260)(775,272)(775,286)(775,293)(775,301)(775,309)(775,316)(775,323)
\Thicklines \path(775,323)(775,330)(775,335)(775,339)(775,342)(775,345)(775,346)(775,349)(775,351)(775,353)(775,354)(775,355)(775,357)(775,359)(775,360)(775,364)(775,368)(775,372)(775,380)(775,389)(775,398)(775,416)(775,435)(776,476)(776,514)(777,554)(778,627)(779,660)(780,688)(782,732)(783,748)(783,755)(784,760)(784,764)(785,766)(785,767)(785,768)(785,768)(785,769)(785,769)(785,769)(785,769)(785,769)(785,769)(786,769)(786,769)(786,769)(786,769)(786,769)(786,769)(786,769)
\Thicklines \path(786,769)(786,769)(786,769)(786,769)(786,768)(786,768)(787,766)(787,765)(787,763)(787,759)(788,753)(788,746)(788,738)(789,728)(789,707)(789,693)(790,680)(790,665)(790,658)(790,650)(790,644)(790,636)(790,629)(790,625)(790,621)(790,619)(790,617)(790,615)(790,613)(790,612)(790,611)(790,608)(790,606)(790,604)(790,602)(790,600)(790,598)(790,596)(790,591)(790,586)(790,578)(790,571)(790,553)(790,534)(790,515)(789,480)
\Thicklines \dashline[-20]{1}(789,480)(789,480)(814,432)(839,387)(858,353)(872,329)(877,320)(880,314)(881,311)(882,311)(882,310)(882,310)(882,309)(882,309)(882,309)(882,309)(882,309)(882,309)(882,309)(882,309)(882,309)(882,309)(882,309)(882,308)(882,308)(882,308)(882,308)(882,308)(882,308)(881,308)(881,308)(881,308)(881,308)(881,308)(881,308)(881,308)(881,309)(880,309)(880,309)(876,313)(865,327)(848,351)(825,387)(800,431)(776,481)(750,539)(727,596)(708,647)(694,693)(685,722)(684,726)
\Thicklines \dashline[-20]{1}(684,726)(684,728)(683,730)(683,730)(683,731)(683,731)(683,731)(683,731)(683,731)(683,731)(683,731)(683,731)(683,732)(683,732)(683,732)(683,732)(683,732)(683,732)(683,732)(683,732)(683,732)(683,731)(684,731)(685,729)(688,720)(700,689)(718,642)(740,588)(764,533)(789,480)
\put(639,480){\raisebox{-1.2pt}{\makebox(0,0){$\Diamond$}}}
\put(679,228){\makebox(0,0){$+$}}
\put(779,191){\raisebox{-1.2pt}{\makebox(0,0){$\Box$}}}
\put(882,308){\makebox(0,0){$\times$}}
\put(0,500){\mbox{${\cal A}_{CP}$}}
\put(740,0){\mbox{${\cal B} / {\cal B}^{SM}$}}
\end{picture}
\caption{The correlation between ${\cal B} / {\cal B}^{SM}$ and
${\cal A}_{CP}$ as $\theta$ moves, and $\phi_{BR} = 0$
(thin solid line), $\pi / 4$ (dotted line), $\pi / 2$ (thick solid line) and
$3 \pi / 4$ (dashed line), where $\tan\theta = A_{BR} / A^N_{SL}$. We set
$\phi_{SL} = 0$. And, for $A_{SL} = - 2 C_7^{eff}$ and $A_{BR} = - 2 C_7^{eff}$,
plotted some marks, $\Diamond$ ($\phi_{BR} = 0$), $+$ ($\phi_{BR} = \pi / 4$),
$\Box$ ($\phi_{BR} = \phi / 2$) and $\times$ 
($\phi_{BR} = 3 \pi / 4$).}\label{fig:correlationBR}
\end{figure}
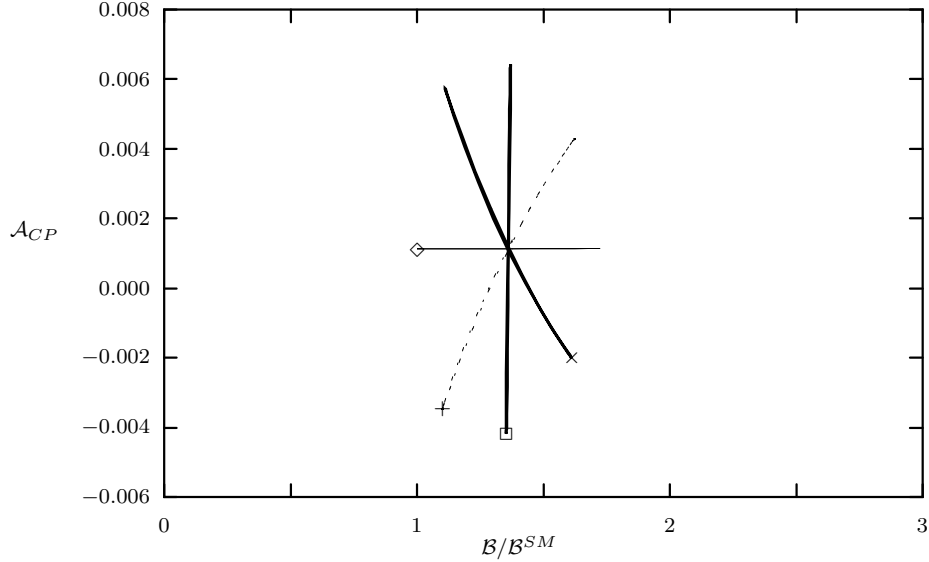
\begin{figure}
\setlength{\unitlength}{0.240900pt}
\begin{picture}(1500,900)(0,0)
\footnotesize
\thicklines \path(242,90)(262,90)
\thicklines \path(1433,90)(1413,90)
\put(220,90){\makebox(0,0)[r]{$-0.006$}}
\thicklines \path(242,199)(262,199)
\thicklines \path(1433,199)(1413,199)
\put(220,199){\makebox(0,0)[r]{$-0.004$}}
\thicklines \path(242,309)(262,309)
\thicklines \path(1433,309)(1413,309)
\put(220,309){\makebox(0,0)[r]{$-0.002$}}
\thicklines \path(242,418)(262,418)
\thicklines \path(1433,418)(1413,418)
\put(220,418){\makebox(0,0)[r]{$0.000$}}
\thicklines \path(242,528)(262,528)
\thicklines \path(1433,528)(1413,528)
\put(220,528){\makebox(0,0)[r]{$0.002$}}
\thicklines \path(242,637)(262,637)
\thicklines \path(1433,637)(1413,637)
\put(220,637){\makebox(0,0)[r]{$0.004$}}
\thicklines \path(242,747)(262,747)
\thicklines \path(1433,747)(1413,747)
\put(220,747){\makebox(0,0)[r]{$0.006$}}
\thicklines \path(242,856)(262,856)
\thicklines \path(1433,856)(1413,856)
\put(220,856){\makebox(0,0)[r]{$0.008$}}
\thicklines \path(242,90)(242,110)
\thicklines \path(242,856)(242,836)
\put(242,45){\makebox(0,0){$0$}}
\thicklines \path(441,90)(441,110)
\thicklines \path(441,856)(441,836)
\thicklines \path(639,90)(639,110)
\thicklines \path(639,856)(639,836)
\put(639,45){\makebox(0,0){$1$}}
\thicklines \path(838,90)(838,110)
\thicklines \path(838,856)(838,836)
\thicklines \path(1036,90)(1036,110)
\thicklines \path(1036,856)(1036,836)
\put(1036,45){\makebox(0,0){$2$}}
\thicklines \path(1235,90)(1235,110)
\thicklines \path(1235,856)(1235,836)
\thicklines \path(1433,90)(1433,110)
\thicklines \path(1433,856)(1433,836)
\put(1433,45){\makebox(0,0){$3$}}
\thicklines \path(242,90)(1433,90)(1433,856)(242,856)(242,90)
\thinlines \path(789,480)(789,480)(752,480)(715,480)(684,480)(661,480)(652,480)(645,480)(642,480)(640,480)(640,480)(639,480)(639,480)(639,480)(639,480)(639,480)(639,480)(639,480)(639,480)(639,480)(639,480)(639,480)(639,480)(639,480)(639,480)(639,480)(639,480)(639,480)(639,480)(639,480)(639,480)(639,480)(640,480)(640,480)(640,480)(640,480)(640,480)(640,480)(640,480)(640,480)(641,480)(641,480)(641,480)(641,480)(641,480)(642,480)(642,480)(643,480)(644,480)(646,480)(649,480)
\thinlines \path(649,480)(657,480)(679,480)(709,480)(742,480)(781,480)(818,481)(852,481)(868,481)(875,481)(883,481)(889,481)(895,481)(898,481)(899,481)(900,481)(902,481)(902,481)(903,481)(903,481)(904,481)(905,481)(905,481)(906,481)(906,481)(906,481)(907,481)(907,481)(908,481)(908,481)(910,481)(911,481)(912,481)(914,481)(915,481)(917,481)(920,481)(921,481)(922,481)(923,481)(924,481)(925,481)(925,481)(926,481)(926,481)(926,481)(926,481)(926,481)(926,481)(926,481)(926,481)
\thinlines \path(926,481)(926,481)(926,481)(926,481)(926,481)(926,481)(926,481)(926,481)(926,481)(926,481)(925,481)(924,481)(922,481)(917,481)(908,481)(887,480)(857,480)(823,480)(789,480)
\thinlines \dashline[-10]{5}(787,491)(787,491)(761,438)(733,377)(711,322)(694,276)(683,242)(680,233)(679,230)(679,229)(679,228)(678,228)(678,228)(678,228)(678,228)(678,228)(678,228)(678,228)(678,228)(678,228)(678,228)(678,228)(679,228)(679,228)(679,229)(681,235)(691,261)(706,304)(727,355)(752,415)(779,471)(804,521)(830,567)(852,603)(869,628)(881,646)(884,650)(886,652)(886,653)(886,653)(886,653)(886,653)(886,653)(886,653)(886,653)(886,653)(886,653)(886,653)(886,653)(886,652)(884,650)
\thinlines \dashline[-10]{5}(884,650)(875,637)(861,615)(839,582)(815,541)(787,491)
\Thicklines \path(783,496)(783,496)(781,423)(780,347)(779,285)(779,258)(779,238)(779,228)(779,223)(779,219)(779,217)(779,215)(779,214)(779,213)(779,212)(779,212)(779,211)(779,211)(779,210)(779,209)(779,209)(779,209)(779,208)(779,208)(779,207)(779,206)(779,206)(779,205)(779,204)(779,203)(779,201)(779,200)(779,199)(779,198)(779,197)(779,196)(779,195)(779,193)(779,193)(779,192)(779,192)(779,191)(779,191)(779,191)(779,191)(779,191)(779,191)(779,191)(779,190)(779,191)(779,191)
\Thicklines \path(779,191)(779,191)(779,191)(779,191)(779,191)(779,191)(779,192)(779,193)(779,194)(780,197)(780,208)(780,221)(781,263)(781,323)(782,391)(782,463)(783,541)(783,612)(784,671)(784,700)(785,722)(785,741)(785,754)(785,764)(786,767)(786,768)(786,769)(786,769)(786,769)(786,770)(786,770)(786,770)(786,770)(786,770)(786,770)(786,770)(786,770)(786,770)(786,770)(786,769)(786,769)(786,769)(786,769)(786,768)(786,768)(786,767)(786,767)(786,766)(786,765)(786,764)(786,762)
\Thicklines \path(786,762)(786,761)(786,760)(786,759)(786,758)(786,758)(786,757)(786,756)(786,756)(786,755)(786,754)(786,753)(786,753)(786,752)(786,751)(786,750)(786,748)(786,744)(786,741)(786,732)(786,723)(786,699)(785,644)(784,580)(783,505)(783,496)
\Thicklines \path(778,492)(778,492)(803,441)(828,394)(849,358)(865,333)(877,315)(880,311)(882,308)(882,308)(882,308)(882,308)(882,308)(882,308)(882,308)(882,308)(882,308)(882,308)(882,308)(881,309)(880,312)(870,326)(855,349)(835,381)(810,425)(784,474)(759,527)(735,586)(714,640)(697,688)(687,720)(684,728)(683,731)(683,731)(683,732)(683,732)(683,732)(683,732)(683,732)(683,732)(683,732)(683,732)(683,732)(683,732)(683,732)(684,731)(686,723)(695,697)(711,654)(730,602)(755,543)
\Thicklines \path(755,543)(778,492)
\put(639,480){\raisebox{-1.2pt}{\makebox(0,0){$\Diamond$}}}
\put(679,228){\makebox(0,0){$+$}}
\put(779,191){\raisebox{-1.2pt}{\makebox(0,0){$\Box$}}}
\put(882,308){\makebox(0,0){$\times$}}
\put(0,500){\mbox{${\cal A}_{CP}$}}
\put(740,0){\mbox{${\cal B} / {\cal B}^{SM}$}}
\end{picture}
\caption{The correlation between ${\cal B} / {\cal B}^{SM}$ and
${\cal A}_{CP}$ as $\theta$ moves, and $\phi_{NL} = 0$
(thin solid line), $\pi / 4$ (thin dotted line), $\pi / 2$ (thick solid line)
and $3 \pi / 4$ (thick solid line), where $\tan\theta = A_{BR} / A^N_{SM}$. We
set $\phi_{NL} \equiv \phi_{SL} = \phi_{BR}$. And, for $A_{SL} = - 2 C_7^{eff}$
and $A_{BR} = - 2 C_7^{eff}$, I plotted some marks, $\Diamond$ 
($\phi_{NL} = 0$), $+$ ($\phi_{NL} = \pi / 4$), $\Box$ ($\phi_{NL} = \phi / 2$)
and $\times$ ($\phi_{NL} = 3 \pi / 4$).}\label{fig:correlationNL}
\end{figure}
\begin{figure}
\setlength{\unitlength}{0.240900pt}
\begin{picture}(1500,900)(0,0)
\footnotesize
\thicklines \path(198,90)(218,90)
\thicklines \path(1433,90)(1413,90)
\put(176,90){\makebox(0,0)[r]{$0.01$}}
\thicklines \path(198,199)(218,199)
\thicklines \path(1433,199)(1413,199)
\put(176,199){\makebox(0,0)[r]{$0.02$}}
\thicklines \path(198,309)(218,309)
\thicklines \path(1433,309)(1413,309)
\put(176,309){\makebox(0,0)[r]{$0.03$}}
\thicklines \path(198,418)(218,418)
\thicklines \path(1433,418)(1413,418)
\put(176,418){\makebox(0,0)[r]{$0.04$}}
\thicklines \path(198,528)(218,528)
\thicklines \path(1433,528)(1413,528)
\put(176,528){\makebox(0,0)[r]{$0.05$}}
\thicklines \path(198,637)(218,637)
\thicklines \path(1433,637)(1413,637)
\put(176,637){\makebox(0,0)[r]{$0.06$}}
\thicklines \path(198,747)(218,747)
\thicklines \path(1433,747)(1413,747)
\put(176,747){\makebox(0,0)[r]{$0.07$}}
\thicklines \path(198,856)(218,856)
\thicklines \path(1433,856)(1413,856)
\put(176,856){\makebox(0,0)[r]{$0.08$}}
\thicklines \path(198,90)(198,110)
\thicklines \path(198,856)(198,836)
\put(198,45){\makebox(0,0){$0$}}
\thicklines \path(404,90)(404,110)
\thicklines \path(404,856)(404,836)
\put(404,45){\makebox(0,0){$0.5$}}
\thicklines \path(610,90)(610,110)
\thicklines \path(610,856)(610,836)
\put(610,45){\makebox(0,0){$1$}}
\thicklines \path(816,90)(816,110)
\thicklines \path(816,856)(816,836)
\put(816,45){\makebox(0,0){$1.5$}}
\thicklines \path(1021,90)(1021,110)
\thicklines \path(1021,856)(1021,836)
\put(1021,45){\makebox(0,0){$2$}}
\thicklines \path(1227,90)(1227,110)
\thicklines \path(1227,856)(1227,836)
\put(1227,45){\makebox(0,0){$2.5$}}
\thicklines \path(1433,90)(1433,110)
\thicklines \path(1433,856)(1433,836)
\put(1433,45){\makebox(0,0){$3$}}
\thicklines \path(198,90)(1433,90)(1433,856)(198,856)(198,90)
\thinlines \path(1433,139)(1347,158)(1212,196)(1096,241)(988,297)(897,361)(816,440)(752,523)(705,608)(686,650)(669,694)(657,733)(649,764)(646,779)(644,792)(643,798)(643,801)(643,803)(643,805)(643,806)(643,808)(643,809)(643,810)(643,811)(643,811)(643,812)(643,813)(643,813)(643,814)(643,815)(643,816)(643,818)(643,819)(643,820)(643,822)(644,823)(644,825)(645,826)(645,828)(646,829)(646,830)(647,832)(648,832)(649,833)(650,834)(650,834)(651,834)(652,835)(653,835)(653,835)(654,835)
\thinlines \path(654,835)(654,835)(655,835)(655,835)(656,835)(656,835)(657,834)(658,834)(660,834)(663,832)(666,831)(673,826)(710,793)(759,746)(826,686)(905,625)(994,568)(1103,512)(1222,463)(1363,416)(1433,397)
\thinlines \dashline[-10]{5}(1433,148)(1317,176)(1181,220)(1054,276)(949,341)(857,421)(786,506)(735,591)(716,632)(700,674)(694,692)(689,709)(685,724)(683,737)(682,743)(682,746)(681,749)(681,751)(681,752)(681,754)(681,755)(681,755)(681,757)(681,757)(681,758)(681,759)(681,760)(681,761)(681,762)(681,764)(681,765)(682,766)(682,768)(682,770)(683,772)(683,773)(684,774)(685,776)(685,777)(687,778)(688,779)(689,780)(690,780)(691,780)(691,780)(692,780)(693,781)(693,781)(694,781)(694,781)(695,781)
\thinlines \dashline[-10]{5}(695,781)(696,781)(696,780)(697,780)(698,780)(699,780)(702,779)(706,777)(709,776)(718,771)(737,757)(790,713)(859,659)(950,597)(1056,538)(1174,484)(1317,433)(1433,400)
\Thicklines \path(1433,189)(1328,219)(1195,269)(1085,325)(987,392)(910,462)(854,532)(832,566)(812,602)(798,632)(790,657)(788,662)(787,668)(786,673)(785,678)(785,680)(784,682)(784,683)(784,685)(784,685)(784,686)(784,686)(784,687)(784,688)(784,689)(784,690)(784,691)(784,691)(785,692)(785,694)(785,696)(785,697)(786,699)(787,701)(787,702)(788,703)(789,704)(790,705)(791,705)(792,706)(793,707)(795,707)(796,708)(796,708)(797,708)(798,708)(798,708)(798,708)(799,708)(800,708)(800,708)
\Thicklines \path(800,708)(800,708)(801,708)(802,708)(802,708)(803,708)(805,708)(806,707)(810,706)(818,704)(826,700)(872,674)(938,633)(1021,586)(1116,539)(1235,489)(1367,444)(1433,425)
\Thicklines \dashline[-20]{1}(1433,240)(1419,244)(1294,290)(1189,340)(1096,398)(1023,457)(969,514)(947,541)(928,569)(914,594)(906,613)(902,622)(901,626)(900,630)(900,632)(900,634)(900,636)(899,638)(899,639)(899,639)(899,639)(899,640)(899,640)(899,641)(899,641)(899,642)(899,643)(899,643)(899,644)(899,644)(899,645)(900,646)(900,647)(900,648)(900,649)(901,652)(902,653)(902,654)(904,655)(905,656)(907,657)(909,658)(910,659)(911,659)(912,659)(913,659)(913,659)(914,659)(915,659)(915,659)(916,659)
\Thicklines \dashline[-20]{1}(916,659)(917,659)(917,659)(918,659)(919,659)(919,659)(920,659)(921,659)(923,659)(927,658)(931,658)(935,656)(954,650)(979,639)(1036,611)(1113,575)(1203,536)(1316,493)(1433,455)
\put(0,500){\mbox{${\cal A}_{CP}$}}
\put(740,0){\mbox{${\cal B} / {\cal B}^{SM}$}}
\end{picture}
\caption{The correlation of ${\cal B} / {\cal B}_0$ and 
${\cal A}_{CP} / {\cal A}_{CP}^{SM}$ as $A_{LL}$ moves for $\phi_{LL} = \pi / 2$
and $\phi_{NL} = 0$ (thin solid line), $\pi / 4$ (dotted line), $\pi / 2$ (thick
solid line) and $3 \pi / 4$ (dashed line). Here the definition of $\phi_{NL}$ is
the same as Figure
\protect{\ref{fig:correlationNL}}.}\label{fig:correlationLLNL}
\end{figure}
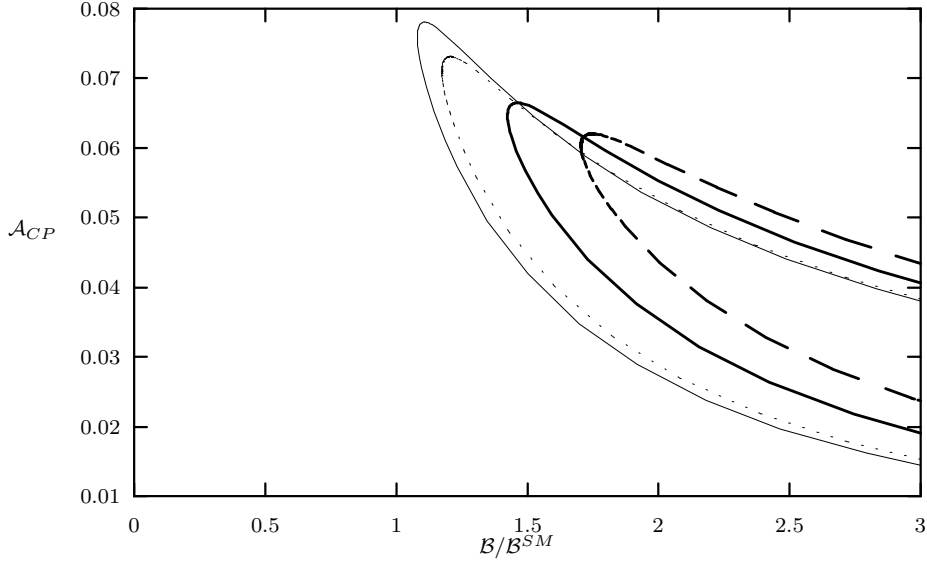

\section{Summary}\label{sec:summary}
I presented the model-independent analysis of the partially integrated CP
asymmetry of the inclusive rare B decay $\bsll$. CP violation is one of the most
interesting topics to search new physics and understand bariogenesis at early
universe, and many researchers has studied this observable through the both of
experimental and theoretical approaches. The process $\bsll$ is experimentally
clean, and there is a possibility that this mode is found by KEKB and PEP-II B
factories. Because $\bsll$ is flavor changing neutral current (FCNC) process, it
is the most sensitive to the various extensions of the standard model (SM). Our
analysis includes the full operator basis, that is, twelve independent four
Fermi operators. In the SM, only three type Wilson coefficients contribute to
$\bsll$, and the partially integrated CP asymmetry has order of $10^{-3}$. We
investigated the correlation of the partially integrated branching ratio and the
partially integrated CP asymmetry, and then can conclude that only $C_{LL}$, the
coefficient of the operator ($\LLop$), can give meaningful contribution to our
process. This cause is the same as the branching ratio\cite{FKMY}, that is, the
large interference between $(B_9 - C_{10})$ and $C_{LL}$. Since
$(B_9 + C_{10}) \ll (B_9 - C_{19})$, the contribution of $C_{LR}$,
the coefficient of operator ($\LRop$), is less than $C_{LL}$. However, the Wilson
coefficients of the other new local interactions beyond SM work only to
suppress the asymmetry, because we assumed there was no new strong phase, and then
they have no interference with the SM interactions. In order to contrast with
the left-right symmetric model, we made $C_{RL}$ and $C_{RR}$ have very small
strong phase, however it changes the size of the CP asymmetry a little. For
$C_{BR}$ and $C_{SL}$, the coefficients of the $\bsg$ operators, although the
asymmetry depends on the weak phase $\phi_{BR}$ of $C_{BR}$ largely, their size
gives little contribution to the asymmetry. Thus, the dependency of two
coefficients is much smaller than that of $A_{LL}$. Note that the branching
ratio also depends on the $\phi_{BR}$.

Our analysis contains the special cases like the MSSM and the 2HDM. In the MSSM,
a special case is $C_{BR} = C_{SL} = 2 C_7$, $C_{LL} = C_9^{eff} - C_{10}$ and
$C_{LR} = C_9^{eff} + C_{10}$. This is just expressed as an example by Figure
\ref{fig:correlationNL}. Therefore, the asymmetry is very suppressed like the
standard model, although the branching ratio can be large or not. 
However, the model has the possibility of the conversion of
sign of $C_{10}$. In this case, Figures \ref{fig:correlationLL} and
\ref{fig:correlationLR} show that the CP asymmetry may be enlarged. Large
contributions to $A_{CP}$ was pointed out by Ref. \cite{LSCPasym}. Figure
\ref{fig:correlationNL} includes the rough character of 2HDM, where a new weak
phase enters into $C_{BR}$ and $C_{SL}$ with the deviation from the SM
prediction for the numerical values of $C_{BR}$, $C_{SL}$, $C_{LL}$ and
$C_{LR}$. When $\sin\phi_{NL}$ is small, the asymmetry is suppressed, however
when $\sin\phi_{NL}$ is close to unit, it change with the sign of
$C_{BR}$\cite{Iltan}. Once the CP asymmetry is measured, we will be able to
constrain the extended models by comparing the data with our numerical analysis. If
we will get the signature of the asymmetry in it, we can conclude that there is
a new $(V - A) \otimes (V - A)$ interaction and / or sizable strong
coupling. Otherwise, the analysis of the present paper cannot constrain us
within some models, so we have to wait the future experiments to get some
informations on the CP from $\bsll$.

I would like to thank to C.S. Kim, T. Yoshikawa, and T. Morozumi who give
suggestion and comments.

\appendix
\section*{Appendix}
\section{Kinematic Functions}\label{app:kinfun}
We list a set of the kinematic functions, which decide the behavior of the
branching ratio and the CP asymmetry for the decay $b \to s l^+ l^-$, and show
the general expression of the direct CP asymmetry. The ratio is shown by
Eq.(\ref{eqn:branching}). We follow Refs.\cite{FKMY,FKY} as the notation. That
is, the functions are given by
\bea
S_1(s) &=& - \frac{4}{s} u(s) \{ 
               s^2 - \frac{1}{3} u(s)^2   
                       - ( m_b^2 - m_s^2 )^2  \}, \nn \\
S_2(s) &=& - 16 u(s) m_b m_s, \nn \\
S_3(s) &=& 4 u(s) ( s + m_b^2 -  m_s^2), \nn \\
S_4(s) &=& 4 u(s) ( s - m_b^2 +  m_s^2), \nn \\
M_1(s) &=& (m_s^2 + m_b^2) S_1(s) + 2 m_b m_s S_2(s),  \nn\\
M_2(s) &=& 2 u(s) ( - \frac{1}{3} u(s)^2  - s^2 + ( m_b^2 - m_s^2)^2 ), \nn \\
M_4(s) &=& m_s^2 S_3(s) + m_b^2 S_4(s)\nn\\
M_6(s) &=& m_b m_s (S_3(s) + S_4(s)),  \nn \\
M_8(s) &=& 2 u(s) ( m_b^2 + m_s^2 -s ) s, \nn \\
M_9(s) &=& 2 u(s) \{- \frac{2}{3} u(s)^2  - 2 ( m_b^2 + m_s^2) s 
      + 2  ( m_b^2 - m_s^2 )^2 \},\label{eqn:kinfunc}
\eea
where we neglected lepton mass.

With the above functions, we can express the partially integrated CP asymmetry
delivered from the matrix element Eq.(\ref{eqn:matrix}), that is,
\beq
	{\cal A}_{CP} 
	\equiv \frac{\int^8_{1 GeV^2} ds (d{\cal N}_{CP}(s)/ds)}
		{\int^8_{1 GeV^2} ds (d{\cal D}_{CP}(s)/ds)}
	\equiv \frac{{\cal N}_{CP}}{{\cal D}_{CP}},
\eeq
where
\bea
\frac{d {\cal N}_{CP}(s)}{d s } &=&	- \frac{1}{{m_b}^8}~{\cal B}_0 [\nn\\ 
			&&\!\!\!\!\!\!\!\!\!\!\!\!\!\!\!\!\!\!\!\!\!\!\!\!\!\!\!\!\!\!\!\!
             S_1(s) ~\{ 2 m_s^2 Im(\lambda_{SL}) Im(B_{SL} A_{SL}^*)
				+ 2 m_b^2 Im(\lambda_{BR}) Im(B_{BR} A_{BR}^*)\} \nn \\
			&&\!\!\!\!\!\!\!\!\!\!\!\!\!\!\!\!\!\!\!\!\!\!\!\!\!\!\!\!\!\!\!\!
			+ S_2(s) ~\{ 2 m_b m_s \left(Im(\lambda_{SL}) Im(A_{SL} B_{BR}^*)
				 + Im(\lambda_{BR}) Im(B_{SL}^* A_{BR}) 
				+ Im(\lambda_{SL}\lambda_{BR}^*) Im(A_{SL} A_{BR}^*)\right)\} \nn \\
           &&\!\!\!\!\!\!\!\!\!\!\!\!\!\!\!\!\!\!\!\!\!\!\!\!\!\!\!\!\!\!\!\!\!\!\!\!\!
			+ S_3(s)~\{ 2 m_s^2 \left(Im(\lambda_{SL}) Im(A_{SL} B_{LL}^*)
				 + Im(\lambda_{LL}) Im(B_{SL}^* A_{LL}) 
				+ Im(\lambda_{SL}\lambda_{LL}^*) Im(A_{SL} A_{LL}^*)\right.\nn\\
			&&\!\!\!\!\!
			\left.+ Im(\lambda_{SL}) Im(A_{SL} B_{LR}^*)
				 + Im(\lambda_{LR}) Im(B_{SL}^* A_{LR}) 
				+ Im(\lambda_{SL}\lambda_{LR}^*) Im(A_{SL} A_{LR}^*)\right)\nn\\
           &&\!\!\!\!\!\!\!\!\!\!\!\!\!\!\!\!\!\!
                   + 2 m_b m_s \left(Im(\lambda_{BR}) Im(A_{BR} B_{RL}^*)
				 + Im(\lambda_{RL}) Im(B_{BR}^* A_{RL}) 
				+ Im(\lambda_{BR}\lambda_{RL}^*) Im(A_{BR} A_{RL}^*)\right.\nn\\
			&&\left.+ Im(\lambda_{BR}) Im(A_{BR} B_{RR}^*)
				 + Im(\lambda_{RR}) Im(B_{BR}^* A_{RR}) 
				+ Im(\lambda_{BR}\lambda_{RR}^*) Im(A_{BR} A_{RR}^*)\right)\}\nn \\
           &&\!\!\!\!\!\!\!\!\!\!\!\!\!\!\!\!\!\!\!\!\!\!\!\!\!\!\!\!\!\!\!\!\!\!\!\!\!
			+ S_4(s)~\{ 2 m_b^2 \left(Im(\lambda_{BR}) Im(A_{BR} B_{LL}^*)
				 + Im(\lambda_{LL}) Im(B_{BR}^* A_{LL}) 
				+ Im(\lambda_{BR}\lambda_{LL}^*) Im(A_{BR} A_{LL}^*)\right.\nn\\
			&&\!\!\!\!\!
			\left. + Im(\lambda_{BR}) Im(A_{BR} B_{LR}^*)
				 + Im(\lambda_{LR}) Im(B_{BR}^* A_{LR}) 
				+ Im(\lambda_{BR}\lambda_{LR}^*) Im(A_{BR} A_{LR}^*)\right)\nn\\
           &&\!\!\!\!\!\!\!\!\!\!\!\!\!\!\!\!\!\!
                   + 2 m_b m_s \left(Im(\lambda_{SL}) Im(A_{SL} B_{RL}^*)
				 + Im(\lambda_{RL}) Im(B_{SL}^* A_{RL}) 
				+ Im(\lambda_{SL}\lambda_{RL}^*) Im(A_{SL} A_{RL}^*)\right.\nn\\
			&&\left.+ Im(\lambda_{SL}) Im(A_{SL} B_{RR}^*)
				 + Im(\lambda_{RR}) Im(B_{SL}^* A_{RR}) 
				+ Im(\lambda_{SL}\lambda_{RR}^*) Im(A_{SL} A_{RR}^*)\right)\} \nn \\ 
             &&\!\!\!\!\!\!\!\!\!\!\!\!\!\!\!\!\!\!\!\!\!\!\!\!\!\!\!\!\!\!\!\!\!\!\!\!\!
			+ M_2(s)~\{ 2 \left( Im(\lambda_{LL}) Im(B_{LL} A_{LL}^*)\right.
				+ Im(\lambda_{LR}) Im(B_{LR} A_{LR}^*)\nn\\
				&&\!\!\!\!\!
             + Im(\lambda_{RL}) Im(B_{RL} A_{RL}^*)
				+ \left.Im(\lambda_{RR}) Im(B_{RR} A_{RR}^*)\right) \} \nn \\
             &&\!\!\!\!\!\!\!\!\!\!\!\!\!\!\!\!\!\!\!\!\!\!\!\!\!\!\!\!\!\!\!\!\!\!\!\!\!
			+ M_6(s)~\{ - 2 \left.( Im(\lambda_{LL}) Im(A_{LL} B_{RL}^*)
				 + Im(\lambda_{RL}) Im(B_{LL}^* A_{RL}) 
				+ Im(\lambda_{LL}\lambda_{RL}^*) Im(A_{LL} A_{RL}^*)\right.\nn\\
			&&\!\!\!\!\! 
				+\left.Im(\lambda_{LR}) Im(A_{LR} B_{RR}^*)
				 + Im(\lambda_{RR}) Im(B_{LR}^* A_{RR}) 
				+ Im(\lambda_{LR}\lambda_{RR}^*) Im(A_{LR} A_{RR}^*)\right) \nn \\
           &&\!\!\!\!\!\!\!\!\!\!\!\!\!\!\!
			+ \left(Im(\lambda_{LRLR}) Im(A_{LRLR} B_{RLLR}^*)
				 + Im(\lambda_{RLLR}) Im(B_{LRLR}^* A_{RLLR})\right. \nn\\
			&&	+ Im(\lambda_{LRLR}\lambda_{RLLR}^*) Im(A_{LRLR} A_{RLLR}^*)\nn\\
           &&\!\!\!\!\!\!\!\!\!\!\!\!\!\!\!
              +Im(\lambda_{LRRL}) Im(A_{LRRL} B_{RLRL}^*)
				 + Im(\lambda_{RLRL}) Im(B_{LRRL}^* A_{RLRL}) \nn\\
			&&\left.+ Im(\lambda_{LRRL}\lambda_{RLRL}^*) Im(A_{LRRL} A_{RLRL}^*)\right)\}
                                \nn \\
             &&\!\!\!\!\!\!\!\!\!\!\!\!\!\!\!\!\!\!\!\!\!\!\!\!\!\!\!\!\!\!\!\!\!\!\!\!\!
				+M_8(s)~\{ 2 \left(Im(\lambda_{LRLR}) Im(B_{LRLR} A_{LRLR}^*)\right.
			+ Im(\lambda_{RLLR}) Im(B_{RLLR} A_{RLLR}^*)\nn\\
			&&\!\!\!\!\!+ Im(\lambda_{LRRL}) Im(B_{LRRL} A_{LRRL}^*)
			\left.+ Im(\lambda_{RLRL}) Im(B_{RLRL} A_{RLRL}^*)\right) \} \nn \\
             &&\!\!\!\!\!\!\!\!\!\!\!\!\!\!\!\!\!\!\!\!\!\!\!\!\!\!\!\!\!\!\!\!\!\!\!\!\!
				+ M_9(s)~\{ 32 Im(\lambda_T) Im(B_T A_T^*)
			+ 128 Im(\lambda_{TE}) Im(B_{TE} A_{TE}^*)\} ],\label{eqn:generalcpnumerator}
\eea
and
\bea
\frac{d {\cal D}_{CP}(s)}{d s } &=&	
		2 \left.\frac{d{\cal B}(s)}{ds}\right|_{C_{XX} \to B_{XX}} + 
	\frac{1}{{m_b}^8}~{\cal B}_0 [\nn\\ 
			&&\!\!\!\!\!\!\!\!\!\!\!\!\!\!\!\!\!\!\!\!\!\!\!\!\!\!\!\!\!\!\!\!
             S_1(s) ~\{ m_s^2 \left(\left|A_{SL}\right|^2 
					+ 2 Re(\lambda_{SL}) Re(B_{SL} A_{SL}^*)\right)
				+  m_b^2 \left(\left|A_{BR}\right|^2 
				+ 2 Re(\lambda_{BR}) Re(B_{BR} A_{BR}^*)\right)\} \nn \\
			&&\!\!\!\!\!\!\!\!\!\!\!\!\!\!\!\!\!\!\!\!\!\!\!\!\!\!\!\!\!\!\!\!
			+ S_2(s) ~\{ 2 m_b m_s \left(Re(\lambda_{SL}) Re(A_{SL} B_{BR}^*)
				 + Re(\lambda_{BR}) Re(B_{SL}^* A_{BR}) 
				+ Re(\lambda_{SL}\lambda_{BR}^*) Re(A_{SL} A_{BR}^*)\right)\} \nn \\
           &&\!\!\!\!\!\!\!\!\!\!\!\!\!\!\!\!\!\!\!\!\!\!\!\!\!\!\!\!\!\!\!\!\!\!\!\!\!
			+ S_3(s)~\{ 2 m_s^2 \left(Re(\lambda_{SL}) Re(A_{SL} B_{LL}^*)
				 + Re(\lambda_{LL}) Re(B_{SL}^* A_{LL}) 
				+ Re(\lambda_{SL}\lambda_{LL}^*) Re(A_{SL} A_{LL}^*)\right.\nn\\
			&&\!\!\!\!\!
			\left.+ Re(\lambda_{SL}) Re(A_{SL} B_{LR}^*)
				 + Re(\lambda_{LR}) Re(B_{SL}^* A_{LR}) 
				+ Re(\lambda_{SL}\lambda_{LR}^*) Re(A_{SL} A_{LR}^*)\right)\nn\\
           &&\!\!\!\!\!\!\!\!\!\!\!\!\!\!\!\!\!\!
                   + 2 m_b m_s \left(Re(\lambda_{BR}) Re(A_{BR} B_{RL}^*)
				 + Re(\lambda_{RL}) Re(B_{BR}^* A_{RL}) 
				+ Re(\lambda_{BR}\lambda_{RL}^*) Re(A_{BR} A_{RL}^*)\right.\nn\\
			&&\left.+ Re(\lambda_{BR}) Re(A_{BR} B_{RR}^*)
				 + Re(\lambda_{RR}) Re(B_{BR}^* A_{RR}) 
				+ Re(\lambda_{BR}\lambda_{RR}^*) Re(A_{BR} A_{RR}^*)\right)\}\nn \\
           &&\!\!\!\!\!\!\!\!\!\!\!\!\!\!\!\!\!\!\!\!\!\!\!\!\!\!\!\!\!\!\!\!\!\!\!\!\!
			+ S_4(s)~\{ 2 m_b^2 \left(Re(\lambda_{BR}) Re(A_{BR} B_{LL}^*)
				 + Re(\lambda_{LL}) Re(B_{BR}^* A_{LL}) 
				+ Re(\lambda_{BR}\lambda_{LL}^*) Re(A_{BR} A_{LL}^*)\right.\nn\\
			&&\!\!\!\!\!
			\left. + Re(\lambda_{BR}) Re(A_{BR} B_{LR}^*)
				 + Re(\lambda_{LR}) Re(B_{BR}^* A_{LR}) 
				+ Re(\lambda_{BR}\lambda_{LR}^*) Re(A_{BR} A_{LR}^*)\right)\nn\\
           &&\!\!\!\!\!\!\!\!\!\!\!\!\!\!\!\!\!\!
                   + 2 m_b m_s \left(Re(\lambda_{SL}) Re(A_{SL} B_{RL}^*)
				 + Re(\lambda_{RL}) Re(B_{SL}^* A_{RL}) 
				+ Re(\lambda_{SL}\lambda_{RL}^*) Re(A_{SL} A_{RL}^*)\right.\nn\\
			&&\left.+ Re(\lambda_{SL}) Re(A_{SL} B_{RR}^*)
				 + Re(\lambda_{RR}) Re(B_{SL}^* A_{RR}) 
				+ Re(\lambda_{SL}\lambda_{RR}^*) Re(A_{SL} A_{RR}^*)\right)\} \nn \\ 
             &&\!\!\!\!\!\!\!\!\!\!\!\!\!\!\!\!\!\!\!\!\!\!\!\!\!\!\!\!\!\!\!\!\!\!\!\!\!
			+ M_2(s)~\{ \left|A_{LL}\right|^2 
				+ 2 Re(\lambda_{LL}) Re(B_{LL} A_{LL}^*)
				+ \left|A_{LR}\right|^2
				+ 2 Re(\lambda_{LR}) Re(B_{LR} A_{LR}^*)\nn\\
				&&\!\!\!\!\!
             + \left|A_{RL}\right|^2 
				+ 2 Re(\lambda_{RL}) Re(B_{RL} A_{RL}^*)
				+ \left|A_{RR}\right|^2
				+ 2 Re(\lambda_{RR}) Re(B_{RR} A_{RR}^*)\} \nn \\
             &&\!\!\!\!\!\!\!\!\!\!\!\!\!\!\!\!\!\!\!\!\!\!\!\!\!\!\!\!\!\!\!\!\!\!\!\!\!
			+ M_6(s)~\{ - 2 \left.( Re(\lambda_{LL}) Re(A_{LL} B_{RL}^*)
				 + Re(\lambda_{RL}) Re(B_{LL}^* A_{RL}) 
				+ Re(\lambda_{LL}\lambda_{RL}^*) Re(A_{LL} A_{RL}^*)\right.\nn\\
			&&\!\!\!\!\! 
				+\left.Re(\lambda_{LR}) Re(A_{LR} B_{RR}^*)
				 + Re(\lambda_{RR}) Re(B_{LR}^* A_{RR}) 
				+ Re(\lambda_{LR}\lambda_{RR}^*) Re(A_{LR} A_{RR}^*)\right) \nn \\
           &&\!\!\!\!\!\!\!\!\!\!\!\!\!\!\!
			+ \left(Re(\lambda_{LRLR}) Re(A_{LRLR} B_{RLLR}^*)
				 + Re(\lambda_{RLLR}) Re(B_{LRLR}^* A_{RLLR})\right. \nn\\
			&&	+ Re(\lambda_{LRLR}\lambda_{RLLR}^*) Re(A_{LRLR} A_{RLLR}^*)\nn\\
           &&\!\!\!\!\!\!\!\!\!\!\!\!\!\!\!
              +Re(\lambda_{LRRL}) Re(A_{LRRL} B_{RLRL}^*)
				 + Re(\lambda_{RLRL}) Re(B_{LRRL}^* A_{RLRL}) \nn\\
			&&\left.+ Re(\lambda_{LRRL}\lambda_{RLRL}^*) Re(A_{LRRL} A_{RLRL}^*)\right)\}
                                \nn \\
             &&\!\!\!\!\!\!\!\!\!\!\!\!\!\!\!\!\!\!\!\!\!\!\!\!\!\!\!\!\!\!\!\!\!\!\!\!\!
				+M_8(s)~\{ \left|A_{LRLR}\right|^2
			+ 2 Re(\lambda_{LRLR}) Re(B_{LRLR} A_{LRLR}^*)
			+ \left|A_{RLLR}\right|^2
			+ 2 Re(\lambda_{RLLR}) Re(B_{RLLR} A_{RLLR}^*)\nn\\
			&&\!\!\!\!\!\!\!\!\!\!\!\!\!\!\!\!+ \left|A_{LRRL}\right|^2 
				+ 2 Re(\lambda_{LRRL}) Re(B_{LRRL} A_{LRRL}^*)
			+ \left|A_{RLRL}\right|^2
			+ 2 Re(\lambda_{RLRL}) Re(B_{RLRL} A_{RLRL}^*)\} \nn \\
             &&\!\!\!\!\!\!\!\!\!\!\!\!\!\!\!\!\!\!\!\!\!\!\!\!\!\!\!\!\!\!\!\!\!\!\!\!\!
				+ M_9(s)~\{ 16 \left(\left|A_T\right|^2 + 2 Re(\lambda_T) Re(B_T A_T^*)\right)
			+ 64 \left(\left|A_{TE}\right|^2 + 2 Re(\lambda_{TE}) Re(B_{TE} A_{TE}^*)\right)\}
				].\label{eqn:generalcpdenomitor}
\eea
The first term $d {\cal B} / ds |_{C_{XX} \to B_{XX}}$ in
Eq.(\ref{eqn:generalcpdenomitor}) is the differential branching ratio
given by Eq.(\ref{eqn:branching}) after replacing all Wilson coefficients
$C_{XX}$ with $B_{XX}$, respectively.

\end{document}